\documentclass[%
 aip,
 rsi,
amsmath,amssymb,
reprint,
floatfix
]{revtex4-1}
\usepackage{graphicx}% Include figure files
\usepackage{float}
\usepackage{dcolumn}% Align table columns on decimal point
\usepackage{bm}% bold math
\usepackage[utf8]{inputenc}
\usepackage{mathptmx}
\usepackage{braket}
\usepackage[english]{babel}
\usepackage{xcolor}% Include colors for document elements
\colorlet{RED}{red}
\colorlet{BLUE}{blue}
\usepackage{dcolumn}% Align table columns on decimal pointhttps://v1.overleaf.com/17481932tmwjwzytcbcq#
\usepackage{bm}% bold math
\usepackage[version=4]{mhchem} % HvD: for general chemical formula like \ce{(TiO2)n}
\usepackage{acronym}
\usepackage{adjustbox}
\usepackage{tikz}
\usepackage{microtype} % fixes many of the overflow hbox errors
\usetikzlibrary{calc,shapes.geometric,decorations.pathmorphing,patterns}

\usepackage[T1]{fontenc}
\usepackage{mathptmx}
\usepackage{tablefootnote}
\usepackage{longtable}
\usepackage{tikzorbital}

\definecolor{background-color}{gray}{0.98}
\usepackage[margin=2.3cm,bmargin=1cm,footnotesep=1cm]{geometry}

\begin{document}

\preprint{AIP/123-QED}

\title[Optimized Virtual Spaces]{Quantum Solvers for Plane-Wave Hamiltonians: Abridging Virtual Spaces Through the Optimization of Pairwise Correlations}
% Force line breaks with \\

\author{Eric J. Bylaska}
   \email{eric.bylaska@pnnl.gov}
   \affiliation{Pacific Northwest National Laboratory, Richland, WA 99352, USA}
\author{Duo Song}
   \email{duo.song@pnnl.gov}
   \affiliation{Pacific Northwest National Laboratory, Richland, WA 99352, USA}
\author{Nicholas P. Bauman}
   \email{nicholas.bauman@pnnl.gov}
   \affiliation{Pacific Northwest National Laboratory, Richland, WA 99352, USA}
\author{Karol Kowalski}
   \email{karol.kowalski@pnnl.gov}
    \affiliation{Pacific Northwest National Laboratory, Richland, WA 99352, USA}
\author{Daniel Claudino}
   \email{claudinodc@ornl.gov}
   \affiliation{Quantum Computing Institute,\ Oak\ Ridge\ National\ Laboratory,\ Oak\ Ridge,\ TN,\ 37831,\ USA \\}
   \affiliation{Computer Science and Mathematics,\ Oak\ Ridge\ National\ Laboratory,\ Oak\ Ridge,\ TN,\ 37831,\ USA \\ }
\author{Travis S. Humble}
   \email{humblets@ornl.gov}
   \affiliation{Quantum Computing Institute,\ Oak\ Ridge\ National\ Laboratory,\ Oak\ Ridge,\ TN,\ 37831,\ USA}
   \affiliation{Computational Sciences and Engineering,\ Oak\ Ridge\ National\ Laboratory,\ Oak\ Ridge,\ TN,\ 37831, USA}

\date{\today}

\begin{abstract}
For many-body methods such as MCSCF and CASSCF, in which the number of one-electron orbitals are optimized and independent of basis set used, there are no problems with using plane-wave basis sets. However, for methods currently used in quantum computing such as select configuration interaction (CI) and coupled cluster (CC) methods, it is necessary to have a virtual space that is able to capture a significant amount of electron-electron correlation in the system.  The virtual orbitals in a pseudopotential plane-wave Hartree--Fock calculation, because of Coulomb repulsion, are often scattering states that interact very weakly with the filled orbitals. As a result, very little correlation energy is captured from them. The use of virtual spaces derived from the one-electron operators have also been tried, and while some correlation is captured, the amount is quite low.  To overcome these limitations, we have been developing new classes of algorithms to define virtual spaces by optimizing orbitals from small pairwise CI Hamiltonians, which we term as correlation optimized virtual orbitals with the abbreviation COVOs.  With these procedures we have been able to derive virtual spaces, containing only a few orbitals, that are able to capture a significant amount of correlation. Besides, using these derived basis sets for quantum computing calculations targeting full CI (FCI) quality-results, they can also be used in other many-body approaches, including CC and M{\o}ller--Plesset perturbation theories, and open up the door to many-body calculations for pseudopotential plane-wave basis set methods. For the H$_2$ molecule, we were able to obtain good agreement with FCI/cc-pVTZ results with just 4 virtual orbitals, for both FCI and quantum simulations.
\end{abstract}


\maketitle

\section{Introduction}
Quantum chemistry is one of the first and most successful scientific applications of digital computers~\cite{mulliken1941hyperconjugation,mulliken1941improved,roothaan1951new,boys1950electronic,parr1950lcao,hall1951molecular,boys1956automatic,nesbet1960ground,allen1962electronic,nesbet1963computer,pople1965approximate,kohn1965self,reeves1966algorithm,pulay1969ab}.  This success has led to a large number of research, open source~\cite{dupuis1989general,stanton1992aces,schmidt1993general,briggs1996real,challacombe2000linear,gygi2008architecture,giannozzi2009quantum,deslippe2012berkeleygw,hutter2014cp2k,gonze2016recent,harrison2016madness,apra2020nwchem}, and commercial codes~\cite{g16,kresse1996efficiency,clark2005first,neese2018software,ADF2001,betteridge2003crystals,werner2012molpro,shao2015advances} (For a larger list quantum chemistry software see~\cite{QCcodes}), which are used on a regular basis by tens of thousands of scientists, engineers, and students from a variety of scientific and engineering domains.  With Moore's law as a backdrop~\cite{moore1965cramming}, the cycle of new machines leading to new algorithms stimulated the field for many decades, and as a consequence, a large number of quantum chemistry methods were developed along with a variety of numerical methods to solve them. However, in recent decades, the maturity and success of these codes coupled with the imminent death of Moore's law~\cite{dubash2005moore,rotman2020we} that made numerical software development much more difficult and less accessible to the average scientist, has resulted in the field having priorities other than just new science, such as porting and optimizing these codes to the next generation of computers~\cite{bylaska2017transitioning,bylaska-knl2017,van2020nwchemex,richard2018developing}, standardization of methods~\cite{wilkins2018nsf,crawford2017molecular}, and marketing~\cite{goldbeck2017scientific,hocquet2017only}. 

With the advent of quantum computing, there is excitement again, and quantum chemists are beginning re-think how they carry out quantum chemistry calculations, in particular very accurate and very expensive instances of systems containing strong electron-electron correlations. This is because it is anticipated that quantum computers with 50-100 qubits will be able to surpass classical digital computers for these types of calculations~\cite{preskill2018quantum}. Quantum computing has thus emerged as an alternative avenue to the continuity of quantum chemistry in the long run~\cite{wasielewski2020exploiting}, but poses several challenges that demand careful consideration in order to eventually mature into a viable replacement for classical computers and large, highly parallelizable HPC clusters.

Present quantum devices are plagued by short coherence times and vulnerability to environment interference, i.e., noise. Albeit quantum algorithms have been developed with proved exactness, such as quantum phase estimation, these are not a viable option in the present/near-term time frame. Therefore, it is desirable to limit the operation of quantum processors to a complementary concerted execution with classical counterparts, whereby each of these components is only in charge of those tasks for which it is more suitable. This has materialized into the variational quantum eigensolver (VQE),\cite{peruzzo2014variational} and other hybrid algorithms. Briefly, this class of algorithms strives to find the lowest eigenvalue of a given observable by assuming the associated quantum state can be accurately represented by a trial wave function and whose parameters are varied according to the Rayleigh-Ritz method (variational principle), with these parameters being updated by the classical computer. The burden on the quantum processor can be further alleviated with strategies such as Trotterization, which in turn introduce other challenges,\cite{DisentangledUCC2019, TrotterUCC}, but which can be successfully exploited in the construction of favorable ansatze, as long as there is not imposed predefined form for the trial wave function. This is at the heart of the ADAPT-VQE.\cite{ADAPT_VQE}    

Most high-levels methods for strongly correlated systems in use today (e.g. full configuration interaction (CI), coupled cluster (CC), Green's function (GF) approaches) are based on second-quantized Hamiltonians, which are written in terms of creation and annihilation operators for fermion orbitals.  These methods are amenable to quantum computers, because fermionic creation and annihilation operators can be readily mapped to qubits through the use of some established transformation, among which Jordan-Wigner~\cite{jw}, Bravyi-Kitaev~\cite{bk}, and binary codes~\cite{binary_code} stand out, where the number of qubits scales with the number of orbitals in the second quantized Hamiltonian. In principle, converting the full many-body electronic Hamiltonian to a second quantized form is exact and popular CC and GF approximations based on this form are very accurate. However, this conversion has a drawback in that it requires the introduction of a basis set, which, for computational cost reasons, needs to be small. Typically, these basis sets are composed of atomic-like orbitals generated with heuristics based on an atom calculation for each kind of atom in the system. An example of this type of basis set is the popular Dunning correlation consistent basis set~\cite{dunning1977gaussian,dunning-cc} in which the atomic orbitals are optimized at the CISD level of theory. While the size of this basis set is small compared to other basis sets used in quantum chemistry, such as plane-waves, it still needs to contain a large number of atomic orbitals to produce a truly accurate result.

Solving relevant chemistry problems analogously to what is classically done with MCSCF or FCI on near-term quantum computers that contain 10's to 100's of noisy qubits~\cite{MSR-Femoco2017}, in which only limited numbers of operations can be performed, is a monumental challenge. One way to reduce the cost of these calculations is to develop new procedures for optimizing basis sets. In this manuscript, a new method is presented for generating a plane-wave derived correlation optimized orbitals basis sets. These derived basis sets can also be used in other many-body approaches, including CC theory and can easily be generalized to work with recently developed Filon Integration Strategy for two-electron integrals in periodic systems~\cite{bylaska2020filon}.  This method is different than other plane-wave derived optimized orbital basis sets~\cite{shirley1996optimal,prendergast2009bloch,chen2011electronic}, in that it is based on optimizing small select CI problems rather then fitting one-electron eigenvalue spectra and band structures.  

The paper is organized as follows. In section~\ref{sec:planewaveH}, a brief description of the second-quantized Hamiltonian and the double unitary CC downfolding method that can be used with the pseudopotential plane-wave method is given, followed by comparisons between restricted Hartree--Fock (RHF) calculations using plane-wave and Gaussian basis sets. Using this framework, CI calculations up to 20 virtual orbitals, generated from plane-wave Hartree-Fock and one-electron Hamiltonians, are shown for the H$_2$ molecule.  The variational quantum eigensolver (VQE) quantum computing algorithms used in this work are described in Section~\ref{sec:VQE}. Section~\ref{sec:newvirtuals} presents
a new class of algorithm for generating a virtual space in which the orbitals are generated by minimizing small pairwise CI Hamiltonians, and a complete set of equations for implementing these optimizations is given in subsections~\ref{sec:one-electron}-\ref{sec:two-electron-matrix}.  Using this new type of virtual space, CI calculations up to 18 virtual orbitals are presented for the ground state energy curve of the H$_2$ molecule are presented in section~\ref{sec:H2results} followed by results using quantum computing simulations in section~\ref{sec:H2QCresults}, and lastly the conclusions are given in section~\ref{sec:conclusions}.

\section{Pseudopotential Plane-Wave Many-Body Hamiltonian}
\label{sec:planewaveH}
The non-relativistic electronic Schr{\"o}dinger eigenvalue equation of quantum chemistry can be written as
\begin{equation}
\label{eqn:schrodinger}
H \ket{\Psi({\bf x}_1,{\bf x}_2,...,{\bf x}_{N_e})} = E \ket{\Psi({\bf x}_1,{\bf x}_2,...,{\bf x}_{N_e})}
\end{equation}
where $H$ is the electronic structure Hamiltonian under the Born--Oppenheimer approximation, and $\ket{\Psi({\bf x}_1,{\bf x}_2,...,{\bf x}_{N_e})}$ is the quantum mechanical wavefunction that is a function of the spatial and spin coordinates of the $N_e$ electrons, ${\bf x}_i = ({\bf r}_i, \sigma_i)$.  When solving this equation the Pauli exclusion principle constraint of particle exchange must be enforced, in which the wavefunction changes sign when the coordinates of two particles, ${\bf x}_i$ and ${\bf x}_j$, are interchanged, i.e. 
\begin{eqnarray}
\label{eqn:antisymmetry}
    && \ket{\Psi({\bf x}_1,{\bf x}_2,...{\bf x}_i,...{\bf x}_j,...,{\bf x}_{N_e})} \nonumber \\
    &=&
    -\ket{\Psi({\bf x}_1,{\bf x}_2,...{\bf x}_j,...{\bf x}_i,...,{\bf x}_{N_e})}.
\end{eqnarray}
 
For the Born--Oppenheimer Hamiltonian, the interaction between the electrons and nuclei are described by the proper potentials $\frac{Ze}{|{\bf r}_i - {\bf R}_A|}$, which for plane-wave solvers can cause trouble with convergence because of the singular behavior at $|{\bf r} - {\bf R}_A|$.  A standard way to remove this issue in plane-wave calculations is to replace these singular potentials by pseudopotentials.  By making this replacement, the Hamiltonian, $H$, in Eq.~\ref{eqn:schrodinger} can be written as
\begin{eqnarray}
    H &=& -\frac{1}{2}\sum_{i=1}^{N_e} \nabla_i^2 \nonumber \\ 
    &+& \sum_{i=1}^{N_e} \sum_{A=1}^{N_A} \left(V^{(A)}_{local}(|{\bf r}_i - {\bf R}_A|) + \sum_{lm} \hat{V}^{(A),lm}_{NL} \right) \nonumber \\
    &+&  \sum_{i=1}^{N_e} \sum_{j>i}^{N_e} \frac{1}{|{\bf r}_i - {\bf r}_j|}
\end{eqnarray}
where the first term is the kinetic energy operator, the second term contains the local and non-local pseudpotentials, $V^{(A)}_{local}$ and $\hat{V}^{(A),(i),lm}_{NL}$, that represent the electron-ion interactions, and the last term is the electron-electron repulsion.

Rather than write the many-electronic Hamiltonian in the traditional Schr{\"o}dinger form, as in the equations above, it is more common today to write it in an alternative representation, known as the second-quantization form.  In this form,  single particle (electron) creation $a_p^\dagger \ket{0} = \ket{1}$ and annihilation $a_p\ket{1} = \ket{0}$ operators are introduced, where the occupation of a specified state $p$ is defined as $\ket{1}$ and $\ket{0}$ for the occupied and unoccupied orbitals respectively. The second-quantized Hamiltonian is written as
    \begin{equation} \label{eqn:secondH}
        H = \sum^{N_{basis}}_{p=1} \sum^{N_{basis}}_{q=1} h_{pq}a_p^\dagger a_q + \frac{1}{2}\sum_{pqrs}
        h_{pqrs}
        a_p^\dagger a_r^\dagger a_s a_q,
    \end{equation}
    \small
    \begin{eqnarray*}
        h_{pq} &=& \int d{\bf x} \phi_p^*({\bf x})\left( -\frac{1}{2} \nabla^2
         \right)\phi_q({\bf x}) \\
        +&&\int d{\bf x} \phi_p^*({\bf x})\left(\sum_{A=1}^{N_A} 
        \left(V^{(A)}_{local}(|{\bf r} -{\bf R}_A|) + \sum_{lm} \hat{V}^{(A),lm}_{NL} \right)
        \right)\phi_q({\bf x}) \\
        h_{pqrs} &=& \int d{\bf x}_1 d{\bf x}_2 \phi_p^*({\bf x}_1)\phi_r^*({\bf x}_2)\frac{1}{|{\bf r}_1 - {\bf r}_2|}\phi_s({\bf x}_2) \phi_q({\bf x}_1)
    \end{eqnarray*}
    \normalsize
where $\phi_p\left({\bf x}\right)$ represent one-electron spin-orbital basis.
A nice feature about this form of the Hamiltonian is that the antisymmetry of wavefunction requirement as given in Eq.~\ref{eqn:antisymmetry} is automatically enforced through the standard fermionic anti-commutation relations $\{a_p,a_q^\dagger\}=\delta_{pq}$ and $\{a_p,a_q\}=\{a_p^\dagger,a_q^\dagger\}=0$. 

In this formulation,  the choice of the one-electron spin-orbital basis is nebulous and requires some care in its choosing in order to obtain accurate results with this type of Hamiltonian.  Typically, in quantum chemistry one uses the filled and virtual orbitals from a Hartree--Fock calculation.  For methods that utilize linear combinations of atomic orbitals (LCAO) as the basis, the size of the basis set and subsequently generated Hartree--Fock orbitals is fairly small.  However, for plane-wave solvers, and other grid based solvers, the size of the basis set is very large and the number of the one- and two-electron integrals in Eq.~\ref{eqn:secondH} will become prohibitive if all possible Hartree--Fock orbitals are used.  

One approach to this problem is to only include virtual orbital up to a certain energy threshold, and another related approach 
is to use the plane-wave derived optimized orbital basis set, e.g. the Shirley approach.  While the number of these orbitals needed to accurately describe eigenvalue spectra over a range of ~100 eV is significantly smaller than the number of plane-waves, it is still significantly larger then the number of orbitals generated by an LCAO method. The reason for this is that the virtual orbitals in a plane-wave Hartree--Fock calculation, because of Coulomb repulsion, are often unbound scattering states that interact very weakly with the filled orbitals.  As a result, very little correlation energy is captured from them. In contrast, LCAO basis methods can only describe bound states, and hence Hartree--Fock calculations in this basis do not generate these types of scattering states.

%Table 1 %%%%%%%%%%%%%%%%%%%%%%%%%%%%%%%%%%%%%%%%%%%%%%%%%%%%%%%%%
\begin{table}[]
    \centering
    \begin{tabular}{c|cccc}
         R(H-H)  & PW FCI      & PW QDK &  PW QDK & CCSD  \\
         (\AA)   & 19 $H_1$ Virt. & DUCC 4  & DUCC 6 & cc-pVTZ\\ \hline
         0.423 & -0.99396 & -0.99113 & -0.99052 & -1.01302\\
         0.529 & -1.10715 & -1.10363 & -1.10440 & -1.11770\\
         0.741 & -1.15340 & -1.14968 & -1.15010 & -1.17010\\
         1.058 & -1.11042 & -1.10768 & -1.10785 & -1.13499\\
         1.588 & -1.01251 & -1.01435 & -1.01417 & -1.05458\\
         2.117 & -0.94070 & -0.94303 & -0.94318 & -1.01464\\
         4.233 & -0.83962 & -0.84025 & -0.84156 & -0.99955
    \end{tabular}
    \caption{Total energies as a function of distance from plane-wave FCI calculations for the H$_2$ molecule 19 $H_1$ virtual orbitals.}
    \label{tab:pwducc}
\end{table}
%%%%%%%%%%%%%%%%%%%%%%%%%%%%%%%%%%%%%%%%%%%%%%%%%%%%%%%%%%%%%%%%

%\vspace{1cm}
\subsection{Many-body downfolding techniques}
\label{sec:DUCC}

%Figure 1 %%%%%%%%%%%%%%%%%%%%%%%%%%%%%%%%%%%%%%%%%%%%%%%%%%%%%%
\begin{figure}[!ht]
    \centering
    \includegraphics[width=\columnwidth]{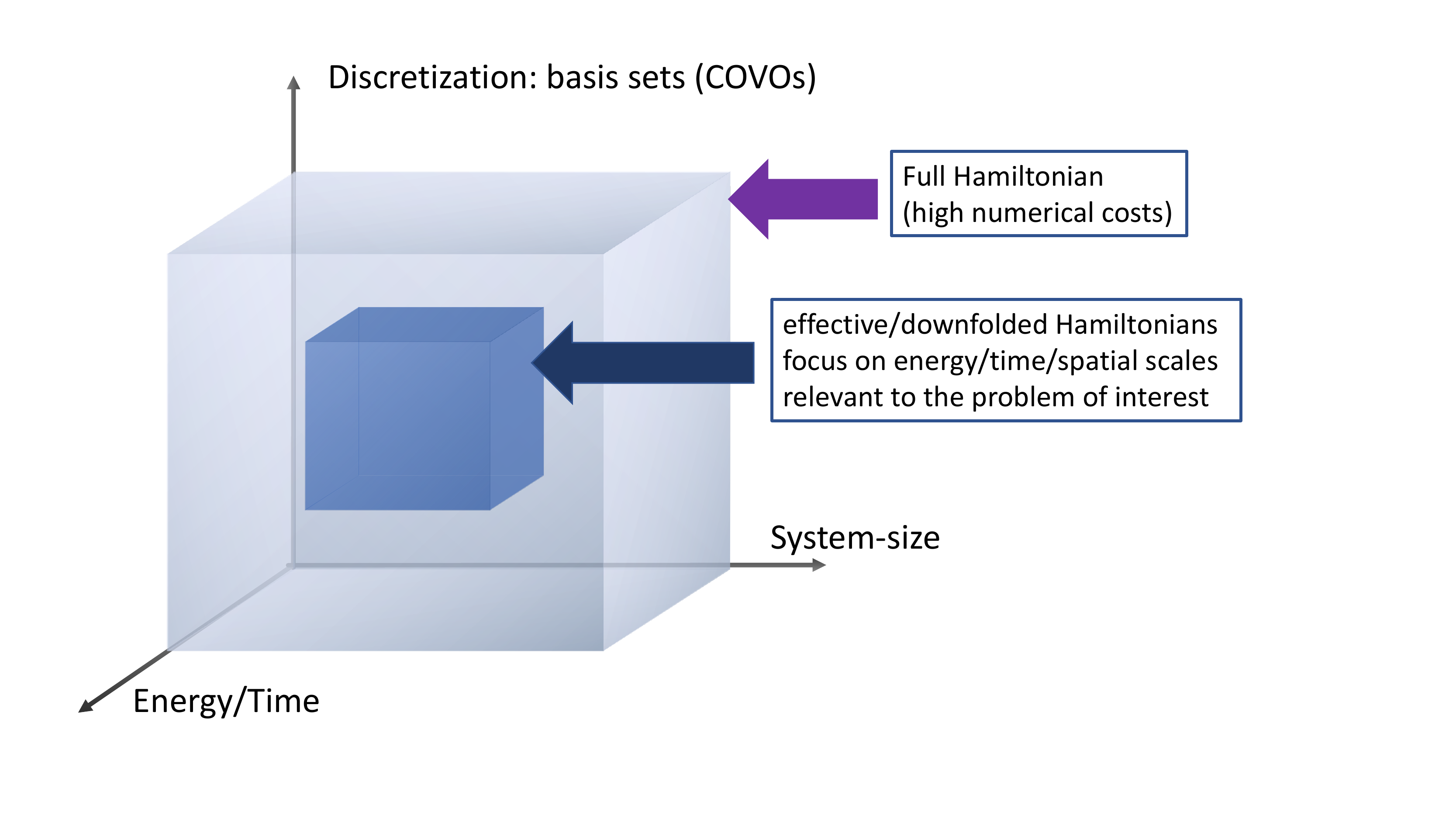}
    \caption{Schematic representation of the dimensionality reduction algorithms considered in this paper: (1) discretization of the many-body problem by employing efficient single-particle basis sets (in this paper, we consider correlation optimized virtual orbitals (COVOs)) and (2)  downfolding techniques based on the double unitary coupled cluster (DUCC) formalism 
    \cite{bauman2019downfolding,kowalski2020sub}
    - in this step the many-body problem is re-represented in a subspace of entire Hilbert space.}
    \label{fig:reduction}
\end{figure}
%%%%%%%%%

One technique for reducing the dimensionality of large plane-wave calculations is to construct effective Hamiltonians that capture correlation effects of the full calculation in manageable active spaces. That way all many-body effects are retained as opposed to simply truncating the orbital space. In Ref.~\cite{bauman2019downfolding} we 
introduced  downfolding techniques, 
which utilize the double unitary CC Ansatz (DUCC)  for exact ground-state wave function $|\Psi\rangle$,
\begin{equation}
	|\Psi\rangle=e^{\sigma_{\rm ext}} e^{\sigma_{\rm int}}|\Phi\rangle\;,
\label{ducc1}
\end{equation}
where $\sigma_{\rm int}$ and $\sigma_{\rm ext}$ are the general type anti-Hermitian operators 
\begin{eqnarray}
\sigma_{\rm int}^{\dagger}&=&-\sigma_{\rm int} \;,
\label{gahint} \\
\sigma_{\rm ext}^{\dagger}&=&-\sigma_{\rm ext}
\label{gahext}
\end{eqnarray}
defined by 
%excitations/de-excitations 
amplitudes defining action 
within and outside of the predefined active space, respectively, i.e.,  the amplitudes defining the $\sigma_{\rm ext}$ operator must carry at least one inactive spin-orbital index whereas all amplitudes defining the $\sigma_{\rm int}$ operator carry active spin-orbital indices only. In Eq.~\ref{ducc1}, $|\Phi\rangle $ designates properly chosen reference function (usually chosen as a Hartree--Fock (HF) Slater determinant). The exactness of the expansion \ref{ducc1}
has been recently discussed in Ref.~\cite{kowalski2020sub} where it was also shown that the standard UCC expansions can provide a basic approximation of the exact $\sigma_{\rm int}$ and $\sigma_{\rm ext}$
operators, i.e.,
\begin{eqnarray}
\sigma_{\rm int} \simeq T_{\rm int} - T_{\rm int}^{\dagger}
\;, \label{uccint} \\
\sigma_{\rm ext} \simeq T_{\rm ext} - T_{\rm ext}^{\dagger}
\;, \label{uccext}
\end{eqnarray}
where $T_{\rm int}$ and $T_{\rm ext}$ are single-reference-type internal and external cluster amplitudes (in the sense defined above).

Using  Ansatz \ref{ducc1} we have shown~\cite{kowalski2020sub} that the  exact energy of the systems can be reproduced by the diagonalization of the effective (or downfolded) Hamiltonian $\overline{H}_{\rm eff}^{\rm (DUCC)}$ in the corresponding active-space:
\begin{equation}
        \overline{H}_{\rm eff}^{\rm (DUCC)} e^{\sigma_{\rm int}} \ket{\Phi} = E e^{\sigma_{\rm int}}\ket{\Phi},
\label{duccstep2}
\end{equation}
where 
\begin{equation}
        \overline{H}_{\rm eff}^{\rm (DUCC)} = (P+Q_{\rm int}) e^{-\sigma_{\rm ext}}H e^{\sigma_{\rm ext}} (P+Q_{\rm int}) \;.
\label{equivducc}
\end{equation}
In Eq.~\ref{equivducc} the $P$ and $Q_{\rm int}$ are the projection operators onto the reference function and all active-space excited Slater determinants (with respect to $|\Phi\rangle$). 

We will discuss the utility of the downfolding techniques in the next section for the ground state calculations of H$_2$. This 
is just one of the two approaches presented in this paper for reducing the dimensionality of the quantum problem (see Fig.\ref{fig:reduction}).

%\vspace{1cm}
\subsection{Results for the $^1\Sigma_g^{+}$ ground state of H$_2$ using virtual space from Hartree-Fock and one-electron Hamiltonians}
\label{sec:h2hfvirt}

%Figure 2 %%%%%%%%%%%%%%%%%%%%%%%%%%%%%%%%%%%%%%%%%%%%%%%%%%%%%%
\begin{figure}[!ht]
    \centering
	\includegraphics[width=\columnwidth]{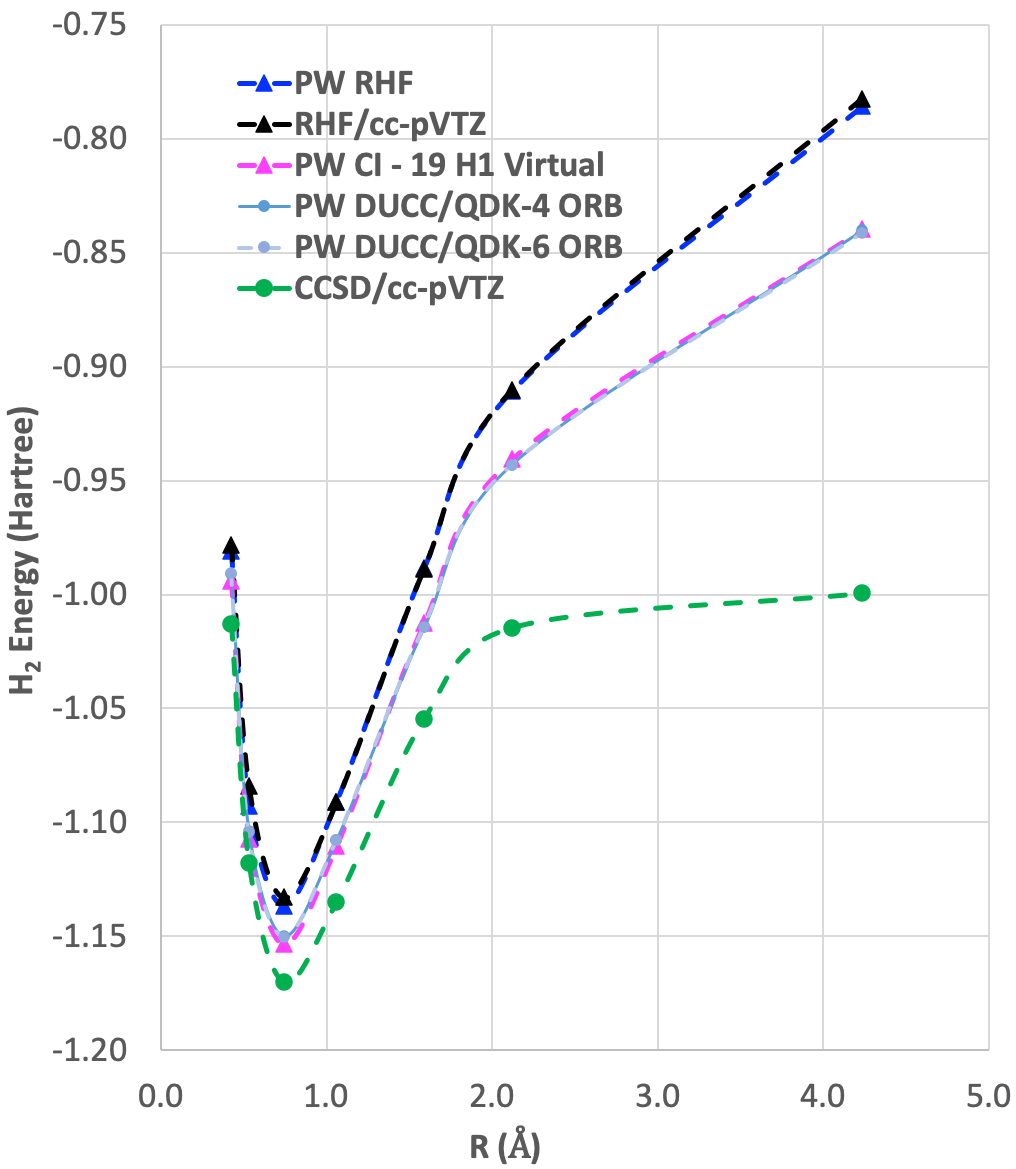}
	\caption{The ground state energy curves for H$_2$ with RHF, CCSD, and DUCC/QDK methods using plane-wave and LCAO Gaussian basis sets. It should be noted that for the a two electron H$_2$ molecule CCSD gives the same answer as FCI.}
	\label{fig:rhfenergy}
\end{figure}
%%%%%%%%%%%%%%%%%%%%%%%%%%%%%%%%%%%%%%%%%%%%%%%%%%%%%%%%%%%%%%%%

The NWChem program package~\cite{kendall2000high,valiev2010nwchem,bylaska2011large,bylaska2017plane,apra2020nwchem} was used for all calculations in this study, except for the FCI calculations, which used the TINYMRCC suite by Ji{\v{r}}{\'\i} Pittner.  The plane-wave calculations used a simple cubic box with L=26$a_0$ and a cutoff energy of 100 Ry.  The RHF and coupled cluster singles and doubles (CCSD) LCAO calculations used the Dunning cc-pVTZ Gaussian basis set. 

%Figure 3 %%%%%%%%%%%%%%%%%%%%%%%%%%%%%%%%%%%%%%%%%%%%%%%%%%%%%%
\begin{figure}[!ht]
    \centering
	\includegraphics[width=\columnwidth]{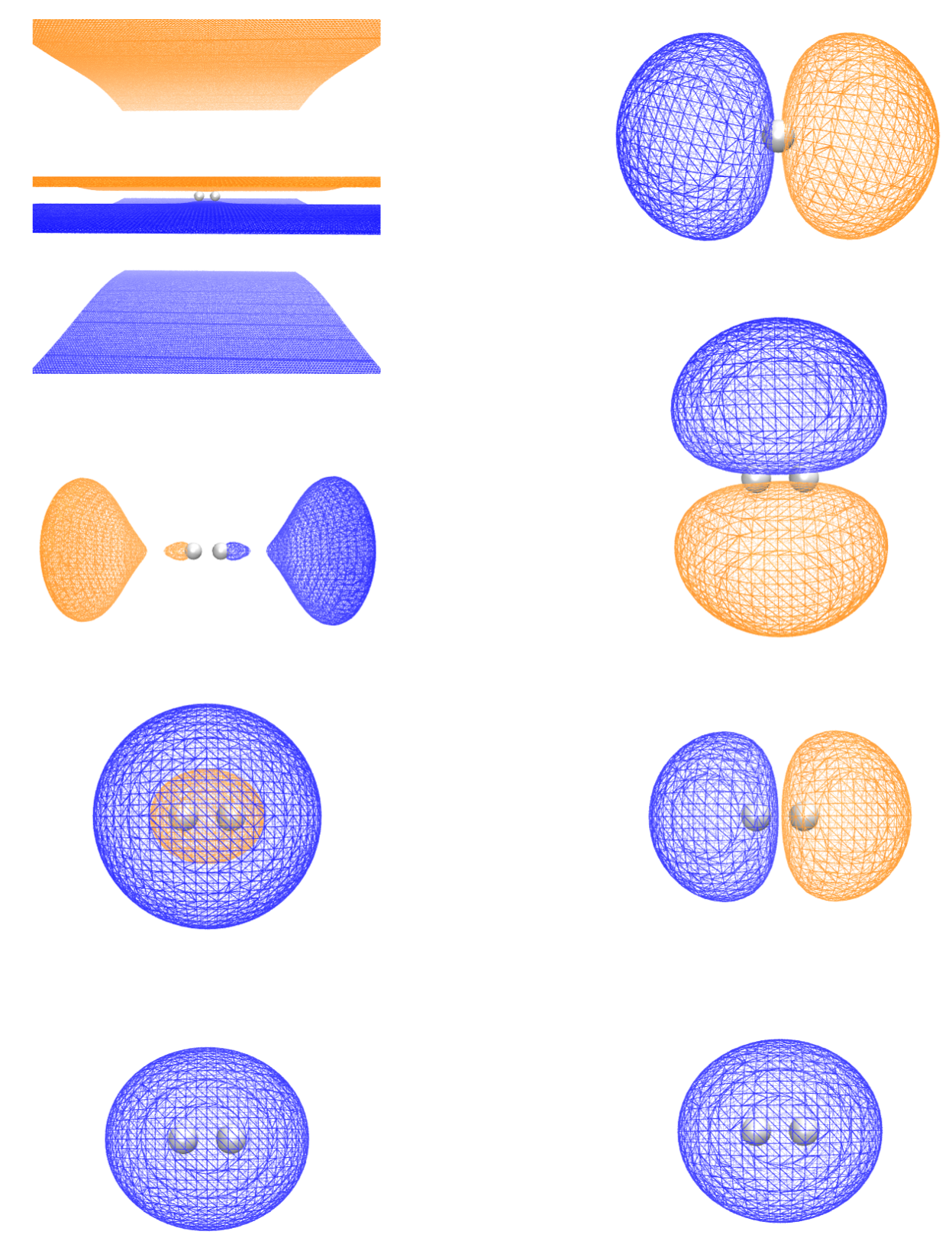}
	\caption{The HOMO and first three LUMOs generated from the straight HF calculation and the $H_1$ Hamiltonian are shown on the left and right panels respectively. The orbitals are displayed in the order of decreasing orbital energy from top to bottom. The isovalues of positive and negative isosurfaces are 0.003117 and -0.003117 for the RHF LUMO 1; 0.01148 and -0.01148 for the RHF LUMO 2; 0.002404 and -0.002404 for the RHF LUMO 3; and 0.03117 and -0.03117 for the others. Notice the orientation of H$_2$ is rotated by 90$^o$ in the bound LUMO 3 relative to the bound LUMO 2.}
	\label{fig:rhforbs}
\end{figure}
%%%%%%%%%%%%%%%%%%%%%%%%%%%%%%%%%%%%%%%%%%%%%%%%%%%%%%%%%%%%%%%%

As can be seen in Fig.~\ref{fig:rhfenergy},  the RHF ground state energy curve of the H$_2$ molecule using plane-wave and LCAO Gaussian basis sets give nearly identical results. However, when we performed plane-wave FCI calculations (not shown) for this system using up to 20 RHF virtual orbitals, the amount of correlation energy calculated was nearly zero (<1.0e-4 Hartree). This result was not surprising since most of the virtual states where scattering states as shown in Fig.~\ref{fig:rhforbs}. 

Instead of using virtual states of the RHF Hamiltonian, virtual states were also generated using the 1-electron part of the RHF Hamiltonian, $H_1$ (i.e. just the kinetic energy and pseudopotential terms).  As shown in Fig.~\ref{fig:rhforbs}, the $H_1$ Hamiltonian generated virtual orbitals that were bound, and looked like the virtual orbitals generated in an LCAO RHF calculation. Using these $H_1$ generated orbitals, we performed plane-wave CI calculations using 19 of these virtual orbitals.  As seen in Fig.~\ref{fig:rhfenergy} and Table~\ref{tab:pwducc},  a significant improvement was seen using these orbitals as they were able to capture a non-zero amount of the correlation energy, however, it was still significantly less than found in LCAO calculations.  In addition, results using quantum phase estimation (QPE) algorithm in the Microsoft QDK package~\cite{svore2018q,low2019q} in which the number of orbitals was reduced to 4 and 6 orbitals using the DUCC method are shown.  These results showed that the DUCC QDK QPE method produces total energies that are within a few milli-Hartrees of the 20 orbital FCI result with only 4 or 6 orbitals. 

%%%%%%%%%%%%%%%%%%%%%%%%%%%%%%%%%%%%%%%%%%%%%%%%%%%%
%\vspace{1cm} %% added this because formatting is acting wierd, need to remove 
\section{Variational Quantum Eigensolver Methods}
\label{sec:VQE}
\par %% background on VQE theory, formulation
The variational quantum eigensolver (VQE) is a method to find the quantum state that minimizes a cost function defined in operator form. \cite{OMalley2016,peruzzo2014variational} This is a hybrid computational approach in which preparation of the quantum circuit is tuned using feedback from classical evaluations of the cost function. Reduction of a given problem to minimization, such as solving for the ground-state energy (lowest energy eigenvalue) of a molecular Hamiltonian, may then rely on the variational principle to affirm that only the true ground state could satisfy the minimum energy. \cite{Kandala2017,mccaskey2019quantum}
\par
Formally, we may consider the problem of solving for the ground-state energy, $E_g$, as
\begin{equation}
\label{eq:vqe}
E_g = \min_{\Psi}\langle \Psi | H | \Psi \rangle
\end{equation}
where $H$ represents the second-quantized Hamiltonian of Eq.~\ref{eqn:secondH} and $\Psi$ labels the electronic configuration. Within the context of quantum computing, the fermionic representations of the Hamiltonian and state are transformed to alternate representations in terms of the spin operators. This transformation recast the molecular Hamiltonian into a representation $H_{P}$ that is defined in terms of the usual Pauli spin operators. Similarly, the electronic state $\Psi$ is represented as a variable unitary operator $\hat{U}(\vec{\theta})$ acting on a fiducial quantum state. This yields the equivalent representation of Eq.~\ref{eq:vqe} as 

\begin{equation}
\label{eq:vqe2}
E_g = \min_{\vec{\theta}}\langle 0 |\hat{U}^\dagger (\vec{\theta}) H_P \hat{U}(\vec{\theta})|0 \rangle
\end{equation}
The second equality is pertinent to the current context as it makes explicit the fact that 1) $H_P$, the Hamiltonian in terms of strings of Pauli operators, relates to $H$ through some transformation that maps fermionic creation and annihilation operators to qubits operators, and 2) the trial wave function emerges from the action of the parameterized unitary operator $\hat{U}(\vec{\theta})$ which builds entanglement, usually starting from an un-entangled wave function, such as Hartree--Fock. For practical considerations, $H$ is transformed into $H_P$ with the Jordan-Wigner transformation~\cite{jw}, but alternatives have been reported in the literature~\cite{bk, BKSF-2018, binary_code}. The classical search for the quantum state that minimizes the energy represents the conventional computing task, while evaluation of the expectation value is performed using the quantum computer. In particular, the quantum state is prepared by executing a quantum circuit, which is expressed formally as a series of unitary operators acting on a well-defined initial state. The action of a specific sequence of unitaries is to prepare a given state that is the subject to the measurements necessary to recover the desired expectation value. 
\par 
In practice, the quantum state that minimizes the energy is unknown and, therefore, a search over possible unitaries is necessary to find the form that minimizes the energy. This variational approach to circuit synthesis underlies the VQE method and an essential choice is the selection of a quantum circuit ansatz which defines the range of unitaries that may be formed to prepare and evaluate a quantum state. For electronic structure calculations, seemingly randomized unitaries may offer advantages for efficient circuit construction, but they lack much of the intuition available from theoretical chemistry~\cite{Kandala2017}. Rather, ansatz circuits derived from unitary coupled cluster theory offer a convenient connection to the expected unitary forms of the minimal quantum state~\cite{RomeroUCCSD}.
%% background on VQE for chemistry
\par 
VQE has been applied previously to recovering the electronic energy from the Hamiltonian presented in Eq.~\ref{eqn:secondH}. The literature provides several examples of usage of VQE for problems of chemical interest, both in terms of simulation and implementation on actual quantum hardware. Given the current limitations faced by present quantum computers, these instances are usually accompanied by strategies that reduce the effective Hilbert space, thus leading to a decrease in the computational expense, such as the use active spaces and natural orbitals~\cite{fno}, as well as downfolding techniques introduced earlier. Another route is to modify the form of the ansatz; an example of this alternative would be the so-called Trotterization, which can be used in conjunction with Hilbert space-reducing techniques.
\par %% extension to ADAPT-VQE, UCSD ansatz
Recently, the principle of VQE was extended to use ansatz circuits that are tailored to computational chemistry applications, and specifically the unitary coupled cluster singles and doubles (UCCSD) ansatz state. Adaptive ansatz construction is attractive because it obeys the underlying complexity of the electronic structure in question, whereas a predefined form for the trial wave function in~\ref{eq:vqe} may fall short of the flexibility necessary for intricate problems. The prime example of this class of algorithms is the ADAPT-VQE, which iteratively assembles a circuit according to the expected energy gain signaled by the gradient with respect to the variational parameters.
\par %% background on performance metrics for A/VQE
An important consideration in the performance of both VQE and ADAPT-VQE is the depth of the ansatz circuit and the time required to construct the optimal variational circuit. For electronic structures dominated by weak correlation, ADAPT-VQE tends to be very economical, adding only operators that make a meaningful contribution towards the lowest eigenvalue in the spectrum of the Hamiltonian in Equation~\ref{eqn:Hmatrix}. On the other hand, the usual UCCSD, by virtue of being defined ahead of time, may contain operators with little impact on the energy, but the classical optimizer will still need to perform a number of calls to the cost function in order to find their best values. Also, the gates originating from these operators, even if they are deemed unimportant because of a small associated parameter, will nevertheless be present in the circuit, adding to its depth. If high accuracy is sought, then ADAPT-VQE may require an ansatz comprised of a large number of operators, which in turn adds to the depth of the underlying circuit. More operators also mean more variational parameters, leading to an onerous optimization process. A more detailed analysis of this trade-off can be found in Grisley et al. ~\cite{ADAPT_VQE} 

%\vspace{1cm}
\section{Algorithm for defining a virtual space with a small CI Hamiltonian}
\label{sec:newvirtuals} 
In this section, we present a downfolding method to define virtual orbitals for expanding the second-quantized Hamiltonian given in Eq.~\ref{eqn:secondH}.  These new types of orbitals
are able to capture significantly more correlation energy than the virtual orbitals coming from Hartree–Fock and one-electron Hamiltonians tested in section~\ref{sec:h2hfvirt}.  The basis of this method is to define a set of virtual orbitals, $\{\psi^{(n)}_e(\mathbf{r})\}$ with $n=1..N_{virtual}$, which we call correlation optimized virtual orbitals or COVOs for short, by optimizing a small select configuration interaction (CI) Hamiltonian with respect to a single virtual orbital, and then the next virtual orbitals in sequence, subject to them being orthonormal to the filled and previously computed virtual orbitals.  The algorithm to calculate these new type of orbitals can be formulated as follows:
\begin{enumerate}
    \item Set $n=1$
    \item Using the ground state one-electron orbital, $\psi_g(\mathbf{r})$ (or ground state orbitals for many electron systems), and the virtual orbital to be optimized, $\psi^{(n)}_e(\mathbf{r})$, generate a CI matrix. 
    \item Calculate the select CI expansion coefficients by diagonalizing the CI matrix.
    \item Using the CI coefficients associated with the lowest eigenvalue, calculate the gradient with respect to the $\psi^{(n)}_e(\mathbf{r})$ then update with a conjugate gradient or similar method while making sure that $\psi^{(n)}_e(\mathbf{r})$ is normalized and orthogonal to $\psi_g(\mathbf{r})$ and $\psi^{(m)}_e(\mathbf{r})$ for $m=1,...,n-1$.
    \item If the gradient is small then $n=n+1$
    \item If $n\leq N_{virtual}$ go to step 2, otherwise finished.
\end{enumerate}

In the case of the H$_2$ molecule, a small CI wavefunction for the 2 electron system composed of 2 one-electron orbitals, $\psi_g(\mathbf{r})$ and $\psi^{(n)}_e(\mathbf{r})$, can be written as a linear combination of 6 determinant wavefunctions, or just 3 determinant wavefunctions for just singlet (or triplet) states. 
\begin{eqnarray*}
\Psi_i[\psi_g(\mathbf{r}),\psi_e(\mathbf{r})] 
&=& c_g^{(i)} \Psi_{g}[\psi_g(\mathbf{r})] \\
&+& c_e^{(i)} \Psi_{e}[\psi_e(\mathbf{r})]   \\
&+& c_m^{(i)} \Psi_{m}[\psi_g(\mathbf{r}),\psi_e(\mathbf{r})] + ...  \\
\end{eqnarray*}
Using this small CI ansatz, the energies of the system can be obtained by diagonalizing the following eigenvalue equation.
\begin{equation*}
\label{3x3eig}
H C_i = E_i S C_i
\end{equation*}
where 
\begin{eqnarray}
H & = & 
\begin{bmatrix}
    \left\langle\Psi_g | H | \Psi_g\right\rangle   & 
    \left\langle\Psi_g | H | \Psi_e\right\rangle   &
    \left\langle\Psi_g | H | \Psi_m\right\rangle   \\
    \left\langle\Psi_e | H | \Psi_g\right\rangle    & 
    \left\langle\Psi_e | H | \Psi_e\right\rangle    &
    \left\langle\Psi_e | H | \Psi_m\right\rangle \\
    \left\langle\Psi_m | H | \Psi_g\right\rangle    & 
    \left\langle\Psi_m | H | \Psi_e\right\rangle    &
    \left\langle\Psi_m | H | \Psi_m\right\rangle \\
\end{bmatrix} \label{eqn:Hmatrix} \\
S & = &
\begin{bmatrix}
    \left\langle\Psi_g  | \Psi_g\right\rangle   & 
    \left\langle\Psi_g  | \Psi_e\right\rangle   &
    \left\langle\Psi_g  | \Psi_m\right\rangle   \\
    \left\langle\Psi_e  | \Psi_g\right\rangle    & 
    \left\langle\Psi_e  | \Psi_e\right\rangle    &
    \left\langle\Psi_e  | \Psi_m\right\rangle \\
    \left\langle\Psi_m  | \Psi_g\right\rangle    & 
    \left\langle\Psi_m  | \Psi_e\right\rangle    &
    \left\langle\Psi_m  | \Psi_m\right\rangle \\
\end{bmatrix} \nonumber \\
C_i & = &
\begin{bmatrix}
    c_{g}^{(i)} \\
    c_{e}^{(i)}  \\ 
    c_{m}^{(i)}
\end{bmatrix} \nonumber
\end{eqnarray}

Note the overlap matrix, $S$, is the identity matrix for orthonormal $\psi_g$ and $\psi_e$. The variation with respect to $\psi_e(\mathbf{r})$ can simply obtained using the following formula.
\begin{eqnarray} \label{eqn:psiegradient}
\cfrac{\delta E_i}{\delta \psi^{*}_e(\mathbf{r})} 
&=& c_{g}^{(i)} \cfrac{\delta \left\langle\Psi_g |H|\Psi_g\right\rangle}{\delta \psi^{*}_e(\mathbf{r})} c_{g}^{(i)}  
+ c_{g}^{(i)} \cfrac{\delta \left\langle\Psi_g |H|\Psi_e\right\rangle}{\delta \psi^{*}_e(\mathbf{r})} c_{e}^{(i)} \nonumber \\  
&+& c_{g}^{(i)} \cfrac{\delta \left\langle\Psi_g |H|\Psi_m\right\rangle}{\delta \psi^{*}_e(\mathbf{r})} c_{m}^{(i)}   
+ c_{e}^{(i)} \cfrac{\delta \left\langle\Psi_e |H|\Psi_g\right\rangle}{\delta \psi^{*}_e(\mathbf{r})} c_{g}^{(i)} \nonumber \\  
&+& c_{e}^{(i)} \cfrac{\delta \left\langle\Psi_e |H|\Psi_e\right\rangle}{\delta \psi^{*}_e(\mathbf{r})} c_{e}^{(i)}   
+ c_{e}^{(i)} \cfrac{\delta \left\langle\Psi_e |H|\Psi_m\right\rangle}{\delta \psi^{*}_e(\mathbf{r})} c_{m}^{(i)} \nonumber \\
&+& c_{m}^{(i)} \cfrac{\delta \left\langle\Psi_m |H|\Psi_g\right\rangle}{\delta \psi^{*}_e(\mathbf{r})} c_{g}^{(i)}   
+ c_{m}^{(i)} \cfrac{\delta \left\langle\Psi_m |H|\Psi_e\right\rangle}{\delta \psi^{*}_e(\mathbf{r})} c_{e}^{(i)} \nonumber \\  
&+& c_{m}^{(i)} \cfrac{\delta \left\langle\Psi_m |H|\Psi_m\right\rangle}{\delta \psi^{*}_e(\mathbf{r})} c_{m}^{(i)}  
\end{eqnarray}

It should be noted that the above formulas can be generalized to work beyond two electron systems by using corresponding orbitals techniques~\cite{king1967corresponding,bylaska2018corresponding}.  The next two subsections, \ref{sec:one-electron}-\ref{sec:two-electron-matrix}, provide formulas that can be used to generate the matrix elements in Eq.~\ref{eqn:Hmatrix} and the gradients with respect to $\psi^{*}_e(\mathbf{r})$ in Eq.~\ref{eqn:psiegradient}.

%\vspace{1cm}
\subsection{One-electron orbitals for two-state Hamiltonian}
\label{sec:one-electron}

The four one-electron spin orbitals of two-state Hamiltonian are
\begin{center}
\begin{tikzpicture}
  \drawLevel[elec = up, pos = {(0.0,-0.6)}, width = 1.0]{lowlabel}
  \drawLevel[elec = no, pos = {(0.0,0.6)}, width = 1.0]{highlabel}
  \drawLevel[elec = down, pos = {(2,-0.6)}, width = 1]{}
  \drawLevel[elec = no, pos = {(2,0.6)}, width = 1]{}
  
  \drawLevel[elec = no, pos = {(4,-0.6)}, width = 1]{}
  \drawLevel[elec = up, pos = {(4,0.6)}, width = 1]{}
  \drawLevel[elec = no, pos = {(6.0,-0.6)}, width = 1]{}
  \drawLevel[elec = down, pos = {(6.0,0.6)}, width = 1]{}

  \node[left] at (left lowlabel) {$g$};
  \node[left] at (left highlabel) {$e$};
  \node[above] at (0.5,1.25) {$\ket{\chi_1}$};
  \node[above] at (2.5,1.25) {$\ket{\chi_2}$};
  \node[above] at (4.5,1.25) {$\ket{\chi_3}$};
  \node[above] at (6.5,1.25) {$\ket{\chi_4}$};
\end{tikzpicture}
\end{center}
\begin{eqnarray*}
    \chi_1(\mathbf{x}) &=& \psi_g(\mathbf{r}) \alpha(s)\\
    \chi_2(\mathbf{x}) &=& \psi_g(\mathbf{r}) \beta(s)\\
    \chi_3(\mathbf{x}) &=& \psi_e(\mathbf{r}) \alpha(s)\\
    \chi_4(\mathbf{x}) &=& \psi_e(\mathbf{r}) \beta(s)
\end{eqnarray*}
where the spatial orbitals and spin functions are orthonormalized.
\begin{equation*}
   \int \psi^{*}_g(\mathbf{r})  \psi_e(\mathbf{r}) d\mathbf{r} =  
   \int \psi^{*}_e(\mathbf{r})  \psi_g(\mathbf{r}) d\mathbf{r} = 0 
\end{equation*}

\begin{equation*}
   \int \psi^{*}_g(\mathbf{r})  \psi_g(\mathbf{r}) d\mathbf{r} = \int \psi^{*}_e(\mathbf{r})  \psi_e(\mathbf{r}) d\mathbf{r} = 1 
\end{equation*}

\begin{equation*}
   \int \alpha^{*}(s)  \beta(s) ds = \int \beta^{*}(s) \alpha(s) ds =  0  
\end{equation*}

\begin{equation*}
   \int \alpha^{*}(s)  \alpha(s) ds =  \int \beta^{*}(s)  \beta(s) ds = 1  
\end{equation*}

%\vspace{1cm}
\subsection{Two-electron orbitals for a two-state Hamiltonian}
\label{sec:two-electron}
For the two state system, there are six two-electron wavefunctions, two of which are singlet, two of which are triplet, and two of which contain a mixture of singlet and triplet character.  These wavefunctions can be written as
\begin{center}
\begin{tikzpicture}
  \drawLevel[elec = {pair}, pos = {(0,-0.6)}, width = 0.75]{lowlabel}
  \drawLevel[elec = no, pos = {(0,0.6)}, width = 0.75]{highlabel}
  \drawLevel[elec = no, pos = {(1.25,-0.6)}, width = 0.75]{}
  \drawLevel[elec = pair, pos = {(1.25,0.6)}, width = 0.75]{}
  
  \drawLevel[elec = up, pos = {(2.5,-0.6)}, width = 0.75]{}
  \drawLevel[elec = down, pos = {(2.5,0.6)}, width = 0.75]{}
  \drawLevel[elec = down, pos = {(3.75,-0.6)}, width = 0.75]{}
  \drawLevel[elec = up, pos = {(3.75,0.6)}, width = 0.75]{}

  \drawLevel[elec = up, pos = {(5,-0.6)}, width = 0.75]{}
  \drawLevel[elec = up, pos = {(5,0.6)}, width = 0.75]{}

  \drawLevel[elec = down, pos = {(6.25,-0.6)}, width = 0.75]{}
  \drawLevel[elec = down, pos = {(6.25,0.6)}, width = 0.75]{}

  \node[left] at (left lowlabel) {$g$};
  \node[left] at (left highlabel) {$e$};
  \node[above] at (0.3,1.25) {$\ket{\chi_1 \chi_2}$};
  \node[above] at (1.55,1.25) {$\ket{\chi_3 \chi_4}$};
  \node[above] at (2.8,1.25) {$\ket{\chi_1 \chi_4}$};
  \node[above] at (4.05,1.25) {$\ket{\chi_2 \chi_3}$};
  \node[above] at (5.3,1.25) {$\ket{\chi_1 \chi_3}$};
  \node[above] at (6.55,1.25) {$\ket{\chi_2 \chi_4}$};
\end{tikzpicture}
\end{center}
\small
\begin{eqnarray*}
\ket{\chi_1 \chi_2} &=& \Psi_g(\mathbf{x}_1,\mathbf{x}_2) \\
&=& \left(\psi_g(\mathbf{r}_1)\psi_g(\mathbf{r}_2)\right)
\frac{1}{\sqrt{2}}\left(\alpha(s_1) \beta(s_2) - \alpha(s_2)\beta(s_1) \right) \\
\ket{\chi_3 \chi_4} &=& \Psi_e(\mathbf{x}_1,\mathbf{x}_2) \\
&=& \left( \psi_e(\mathbf{r}_1)\psi_e(\mathbf{r}_2) \right)
\frac{1}{\sqrt{2}}\left(\alpha(s_1) \beta(s_2) - \alpha(s_2)\beta(s_1) \right) \\
\ket{\chi_1 \chi_4} &=& \Psi_a(\mathbf{x}_1,\mathbf{x}_2) \\
&=& \frac{1}{\sqrt{2}} (\psi_g(\mathbf{r}_1)\alpha(s_1)\psi_e(\mathbf{r}_2)\beta(s_2) \;\;\;\;\;\;\;\;\;\; \;\;\;\;\;\;\;\;\;\; \;\;\;\;\;\;\;\;\;\;\\
& &\;\;\;\;\;\; - \psi_e(\mathbf{r}_1)\beta(s_1)\psi_g(\mathbf{r}_2)\alpha(s_2) ) \\
\ket{\chi_2 \chi_3} &=& \Psi_b(\mathbf{x}_1,\mathbf{x}_2) \\ 
&=& \frac{1}{\sqrt{2}}(\psi_g(\mathbf{r}_1)\beta(s_1)\psi_e(\mathbf{r}_2)\alpha(s_2) \;\;\;\;\;\;\;\;\;\; \;\;\;\;\;\;\;\;\;\; \;\;\;\;\;\;\;\;\;\;\\
& & \;\;\;\;\;\; - \psi_e(\mathbf{r}_1)\alpha(s_1)\psi_g(\mathbf{r}_2)\beta(s_2) ) \\
\ket{\chi_1 \chi_3} &=& \Psi_u(\mathbf{x}_1,\mathbf{x}_2) \\
&=& \frac{1}{\sqrt{2}}\left(\psi_g(\mathbf{r}_1)\psi_e(\mathbf{r}_2) - \psi_e(\mathbf{r}_1)\psi_g(\mathbf{r}_2) \right) \left(\alpha(s_1) \alpha(s_2)\right) \\
\ket{\chi_2 \chi_4} &=& \Psi_d(\mathbf{x}_1,\mathbf{x}_2) \\
&=& \frac{1}{\sqrt{2}} \left(\psi_g(\mathbf{r}_1) \psi_e(\mathbf{r}_2) - \psi_e(\mathbf{r}_1)\psi_g(\mathbf{r}_2) \right) \left(\beta(s_1) \beta(s_2)\right)
\end{eqnarray*}
\normalsize

Note that $\Psi_a$ and $\Psi_b$ cannot be written as a product of a spatial wavefunction times a spin-function.  Moreover, these functions are not eigenfunctions of the spin operators $S^2$ and $S_z$, and as a result these determinants contain both singlet and triplet components.  However, if we take linear combinations of them we can get two new wavefunctions that are separable in spatial and spin functions, and at the same time being eigenfunctions of $S^2$ and $S_z$.
\begin{eqnarray*}
\Psi_m &=& \Psi_{a-b} = \frac{1}{\sqrt{2}}\left(\ket{\chi_1 \chi_4} -  (\ket{\chi_2 \chi_3} \right) \\
&=& \frac{1}{\sqrt{2}}\left(\Psi_a(\mathbf{x}_1,\mathbf{x}_2)- \Psi_b(\mathbf{x}_1,\mathbf{x}_2) \right)\\
&=& \frac{1}{\sqrt{2}}\left(\psi_g(\mathbf{r}_1)\psi_e(\mathbf{r}_2)+\psi_e(\mathbf{r}_1)\psi_g(\mathbf{r}_2)\right)\\
&\times&\frac{1}{\sqrt{2}}\left(\alpha(s_1)\beta(s_2)-\beta(s_1)\alpha(s_2)\right)
\end{eqnarray*}
\begin{eqnarray*}
\Psi_p &=& \Psi_{a+b} = \frac{1}{\sqrt{2}}\left(\ket{\chi_1 \chi_4} +  (\ket{\chi_2 \chi_3} \right) \\
&=& \frac{1}{\sqrt{2}}\left(\Psi_a(\mathbf{x}_1,\mathbf{x}_2)+ \Psi_b(\mathbf{x}_1,\mathbf{x}_2) \right)\\
&=& \frac{1}{\sqrt{2}}\left(\psi_g(\mathbf{r}_1)\psi_e(\mathbf{r}_2)-\psi_e(\mathbf{r}_1)\psi_g(\mathbf{r}_2)\right) \\ &\times&\frac{1}{\sqrt{2}}\left(\alpha(s_1)\beta(s_2)+\beta(s_1)\alpha(s_2)\right)
\end{eqnarray*}

%\vspace{1cm}
\subsection{Matrix elements from the one-electron operators}
\label{sec:one-electron-matrix}

The $H_1$ operator for H$_2$ molecule is
\begin{equation*}
    H_1 = h(\mathbf{r}_1) + h(\mathbf{r}_2)
\end{equation*}

where $h(r)$ is a function/operator of the coordinate $r$, i.e.

\begin{equation*}
    h(\mathbf{r}) = -\frac{1}{2} \nabla^2_\mathbf{r} +   \sum_{A=1}^{N_A} \left(V^{(A)}_{local}(|{\bf r} -{\bf R}_A|) + \sum_{lm} \hat{V}^{(A),lm}_{NL} \right)
\end{equation*}
\small
\begin{eqnarray*}
    \bra{\Psi_g} H_1 \ket{\Psi_g}
    &=& 2 \int \psi^{*}_g(\mathbf{r}) h(\mathbf{r}) \psi_g(\mathbf{r}) d\mathbf{r} \\
    \bra{\Psi_g} H_1 \ket{\Psi_e} &=&0\\
    \bra{\Psi_e} H_1 \ket{\Psi_e} &=& 2 \int \psi^{*}_e(\mathbf{r}) h(\mathbf{r}) \psi_e(\mathbf{r}) d\mathbf{r} \\
     \bra{\Psi_m} H_1 \ket{\Psi_m} &=& \int \psi^{*}_g(\mathbf{r}) h(\mathbf{r}) \psi_g(\mathbf{r}) d\mathbf{r} + \int \psi^{*}_e(\mathbf{r}) h(\mathbf{r}) \psi_e(\mathbf{r}) d\mathbf{r}\\
    \bra{\Psi_g} H_1 \ket{\Psi_m} &=&\sqrt{2}\int \psi^{*}_g(\mathbf{r}) h(\mathbf{r}) \psi_e(\mathbf{r}) d\mathbf{r}\\
    \bra{\Psi_m} H_1 \ket{\Psi_g} &=&\sqrt{2}\int \psi^{*}_e(\mathbf{r}) h(\mathbf{r}) \psi_g(\mathbf{r}) d\mathbf{r}\\
    \bra{\Psi_e} H_1 \ket{\Psi_m} &=&\sqrt{2}\int \psi^{*}_e(\mathbf{r}) h(\mathbf{r}) \psi_g(\mathbf{r}) d\mathbf{r} \\
    \bra{\Psi_m} H_1 \ket{\Psi_e} &=&\sqrt{2}\int \psi^{*}_g(\mathbf{r}) h(\mathbf{r}) \psi_e(\mathbf{r}) d\mathbf{r} 
\end{eqnarray*}
\normalsize
\begin{eqnarray*}
    \cfrac{\delta \bra{\Psi_g} H_1 \ket{\Psi_g}}{\delta \psi_e^{*}(\mathbf{r})} &=& 0\\
    \cfrac{\delta \bra{\Psi_g} H_1 \ket{\Psi_e}}{\delta \psi_e^{*}(\mathbf{r})} &=& 0\\
    \cfrac{\delta \bra{\Psi_e} H_1 \ket{\Psi_e}}{\delta \psi_e^{*}(\mathbf{r})} &=& 2 h(\mathbf{r}) \psi_e(\mathbf{r}) \\
   \cfrac{\delta \bra{\Psi_m} H_1 \ket{\Psi_m}}{\delta \psi_e^{*}(\mathbf{r})} &=&  h(\mathbf{r}) \psi_e(\mathbf{r})\\
   \cfrac{\delta \bra{\Psi_g} H_1 \ket{\Psi_m}}{\delta \psi_e^{*}(\mathbf{r})} &=& 0\\
   \cfrac{\delta \bra{\Psi_m} H_1 \ket{\Psi_g}}{\delta \psi_e^{*}(\mathbf{r})} &=& \sqrt{2}h(\mathbf{r}) \psi_g(\mathbf{r}) \\
   \cfrac{\delta \bra{\Psi_e} H_1 \ket{\Psi_m}}{\delta \psi_e^{*}(\mathbf{r})} &=& \sqrt{2}h(\mathbf{r}) \psi_g(\mathbf{r})\\
   \cfrac{\delta \bra{\Psi_m} H_1 \ket{\Psi_e}}{\delta \psi_e^{*}(\mathbf{r})} &=& 0
\end{eqnarray*}

%\vspace{1cm}
\subsection{Matrix elements from the two-electron operators}
\label{sec:two-electron-matrix}

The $H_2$ operator for H$_2$ molecule is
\begin{equation*}
    H_2 = \frac{1}{r_{12}} = \frac{1}{|\mathbf{r}_1 - \mathbf{r}_2|}
\end{equation*}
\small
\begin{eqnarray*}
    \bra{\Psi_g} H_2 \ket{\Psi_g} &=& \iint \psi^{*}_g(\mathbf{r}) \psi_g(\mathbf{r}) \frac{1}{|\mathbf{r} - \mathbf{r}'|} \psi^{*}_g(\mathbf{r}') \psi_g(\mathbf{r}') d\mathbf{r} d\mathbf{r}' \\
    \bra{\Psi_g} H_2 \ket{\Psi_e} &=& \iint \psi^{*}_g(\mathbf{r}) \psi_e(\mathbf{r}) 
    \frac{1}{|\mathbf{r} - \mathbf{r}'|} \psi^{*}_g(\mathbf{r}') \psi_e(\mathbf{r}') d\mathbf{r} d\mathbf{r}'\\
    \bra{\Psi_e} H_2 \ket{\Psi_g} &=& 2 \left[ \int \frac{\psi^{*}_e(\mathbf{r}') \psi_g(\mathbf{r}')}{|\mathbf{r}-\mathbf{r}'|} d\mathbf{r}' \right] \psi_g(\mathbf{r})\\
    \bra{\Psi_e} H_2 \ket{\Psi_e} &=& \iint \psi^{*}_e(\mathbf{r}) \psi_e(\mathbf{r}) 
    \frac{1}{|\mathbf{r} - \mathbf{r}'|} \psi^{*}_e(\mathbf{r}') \psi_e(\mathbf{r}') d\mathbf{r} d\mathbf{r}' \\
     \bra{\Psi_m} H_2 \ket{\Psi_m} &=& \Big[\iint \psi^{*}_e (\mathbf{r})  \psi_e(\mathbf{r}) 
        \frac{1}{|\mathbf{r} - \mathbf{r}'|} 
        \psi^{*}_g(\mathbf{r}') \psi_g(\mathbf{r}') d\mathbf{r} d\mathbf{r}'\\
     & & + \iint \psi^{*}_e (\mathbf{r}) \psi_g(\mathbf{r}) 
         \frac{1}{|\mathbf{r} - \mathbf{r}'|}
         \psi^{*}_g(\mathbf{r}') \psi_e(\mathbf{r}') d\mathbf{r} d\mathbf{r}'\Big]\\
    \bra{\Psi_g} H_2 \ket{\Psi_m} &=& \sqrt{2}\iint \psi^{*}_g(\mathbf{r}) \psi_g(\mathbf{r}) 
    \frac{1}{|\mathbf{r} - \mathbf{r}'|} \psi^{*}_g(\mathbf{r}') \psi_e(\mathbf{r}') d\mathbf{r} d\mathbf{r}'\\
    \bra{\Psi_m} H_2 \ket{\Psi_g} &=& \sqrt{2}\iint \psi^{*}_e(\mathbf{r}) \psi_g(\mathbf{r}) 
    \frac{1}{|\mathbf{r} - \mathbf{r}'|} \psi^{*}_g(\mathbf{r}') \psi_g(\mathbf{r}') d\mathbf{r} d\mathbf{r}'\\
    \bra{\Psi_e} H_2 \ket{\Psi_m} &=& \sqrt{2}\iint \psi^{*}_e(\mathbf{r}) \psi_e(\mathbf{r}) 
    \frac{1}{|\mathbf{r} - \mathbf{r}'|} \psi^{*}_e(\mathbf{r}') \psi_g(\mathbf{r}') d\mathbf{r} d\mathbf{r}'\\
    \bra{\Psi_m} H_2 \ket{\Psi_e} &=& \sqrt{2}\iint \psi^{*}_g(\mathbf{r}) \psi_e(\mathbf{r}) 
    \frac{1}{|\mathbf{r} - \mathbf{r}'|} \psi^{*}_e(\mathbf{r}') \psi_e(\mathbf{r}') d\mathbf{r} d\mathbf{r}'
\end{eqnarray*}
\normalsize
\begin{eqnarray*}
    \cfrac{\delta \bra{\Psi_g} H_2 \ket{\Psi_g}}{\delta \psi_e^{*}(\mathbf{r})} &=& 0 \\
    \cfrac{\delta \bra{\Psi_g} H_2 \ket{\Psi_e}}{\delta \psi_e^{*}(\mathbf{r})} &=& 0 \\
    \cfrac{\delta \bra{\Psi_e} H_2 \ket{\Psi_g}}{\delta \psi_e^{*}(\mathbf{r})} &=& 2 \left[ \int \frac{\psi^{*}_e(\mathbf{r}') \psi_g(\mathbf{r}')}{|\mathbf{r}-\mathbf{r}'|} d\mathbf{r}' \right] \psi_g(\mathbf{r}) \\
    \cfrac{\delta \bra{\Psi_e} H_2 \ket{\Psi_e}}{\delta \psi_e^{*}(\mathbf{r})} &=&  2 \left[ \int \frac{\psi^{*}_e(\mathbf{r}') \psi_e(\mathbf{r}')}{|\mathbf{r}-\mathbf{r}'|} d\mathbf{r}' \right] \psi_e(\mathbf{r}) \\
   \cfrac{\delta \bra{\Psi_m} H_2 \ket{\Psi_m}}{\delta \psi_e^{*}(\mathbf{r})} &=&  \left[ \int \frac{\psi^{*}_g(\mathbf{r}') \psi_g(\mathbf{r}')}{|\mathbf{r}-\mathbf{r}'|} d\mathbf{r}' \right] \psi_e(\mathbf{r}) \\ 
   &+& \left[ \int \frac{\psi^{*}_g(\mathbf{r}') \psi_e(\mathbf{r}')}{|\mathbf{r}-\mathbf{r}'|} d\mathbf{r}' \right] \psi_g(\mathbf{r}) \\
   \cfrac{\delta \bra{\Psi_g} H_2 \ket{\Psi_m}}{\delta \psi_e^{*}(\mathbf{r})} &=& 0 \\
   \cfrac{\delta \bra{\Psi_m} H_2 \ket{\Psi_g}}{\delta \psi_e^{*}(\mathbf{r})} &=& \sqrt{2} \left[ \int \frac{\psi^{*}_g(\mathbf{r}') \psi_g(\mathbf{r}')}{|\mathbf{r}-\mathbf{r}'|} d\mathbf{r}' \right] \psi_g(\mathbf{r}) \\
   \cfrac{\delta \bra{\Psi_e} H_2 \ket{\Psi_m}}{\delta \psi_e^{*}(\mathbf{r})} &=& \sqrt{2}\Big( \left[ \int \frac{\psi^{*}_e(\mathbf{r}') \psi_g(\mathbf{r}')}{|\mathbf{r}-\mathbf{r}'|} d\mathbf{r}' \right] \psi_e(\mathbf{r}) \\ 
   &+& \left[ \int \frac{\psi^{*}_e(\mathbf{r}') \psi_e(\mathbf{r}')}{|\mathbf{r}-\mathbf{r}'|} d\mathbf{r}' \right] \psi_g(\mathbf{r})\Big)\\
   \cfrac{\delta \bra{\Psi_m} H_2 \ket{\Psi_e}}{\delta \psi_e^{*}(\mathbf{r})} &=& \sqrt{2} \left[ \int \frac{\psi^{*}_g(\mathbf{r}') \psi_e(\mathbf{r}')}{|\mathbf{r}-\mathbf{r}'|} d\mathbf{r}' \right] \psi_e(\mathbf{r})
\end{eqnarray*}

%\vspace{1cm}
\section{Results for $^1\Sigma_{ \MakeLowercase{g} }^{+}$ ground state of the H$_2$ molecule using correlation optimized virtual orbitals (COVO\MakeLowercase{s})}
\label{sec:H2results}

%Table 2 %%%%%%%%%%%%%%%%%%%%%%%%%%%%%%%%%%%%%%%%%%%%%%%%%%%%%%%%%
\begin{table}[]
    \centering
    \scriptsize
    \begin{tabular}{c|cccccc}
         $R$(H-H) & PW FCI & PW FCI & PW VQE & PW FCI  & PW FCI & PW FCI  \\
         (\AA)& 1 COVO  & 4 COVOs & 4 COVOs & 8 COVOs  & 12 COVOs & 18 COVOs\\\hline
         0.60 & -1.13749 & -1.15729 & -1.15728 & -1.15902 & -1.16028 & -1.16089\\
         0.70 & -1.15321 & -1.17179 & -1.17178 & -1.17353 & -1.17467 & -1.17525\\
         0.80 & -1.15128 & -1.16858 & -1.16857 & -1.17033 & -1.17136 & -1.17192\\
         0.90 & -1.14124 & -1.15726 & -1.15724 & -1.15903 & -1.15995 & -1.16049\\
         1.00 & -1.12742 & -1.14216 & -1.14213 & -1.14399 & -1.14478 & -1.14533\\
         1.50 & -1.05311 & -1.06195 & -1.06195 & -1.06473 & -1.06516 & -1.06564\\
         2.00 & -1.00793 & -1.01225 & -1.01220 & -1.01868 & -1.01916 & -1.01945\\
         2.50 & -0.98862 & -1.00150 & -1.00150 & -1.00195 & -1.00228 & -1.00301\\
         3.00 & -0.98137 & -0.99704 & -0.99701 & -0.99737 & -0.99789 & -0.99872\\
         3.50 & -0.97883 & -0.99573 & -0.99570 & -0.99629 & -0.99698 & -0.99766\\
         4.00 & -0.97810 & -0.99613 & -0.99611 & -0.99614 & -0.99693 & -0.99736\\
         4.50 & -0.97817 & -0.99609 & -0.99608 & -0.99609 & -0.99722 & -0.99729\\
         5.00 & -0.97845 & -0.99604 & -0.99598 & -0.99603 & -0.99716 & -0.99727\\
         6.00 & -0.97906 & -0.99597 & -0.99596 & -0.99597 & -0.99705 & -0.99719\\
         7.00 & -0.97928 & -0.99596 & -0.99596 & -0.99596 & -0.99703 & -0.99717
    \end{tabular}
    \caption{Total energies as a function of distance for the H$_2$ molecule from plane-wave FCI calculations with 1, 4, 8, 12, and 18 COVOs and ADAPT-VQE simulations with 4 COVOs.}
    \label{tab:pwfci}
\end{table}
%%%%%%%%%%%%%%%%%%%%%%%%%%%%%%%%%%%%%%%%%%%%%%%%%%%%%%%%%%%%%%%%
%Figure 4 %%%%%%%%%%%%%%%%%%%%%%%%%%%%%%%%%%%%%%%%%%%%%%%%%%%%%%
\begin{figure}[!ht]
    \centering
	\includegraphics[width=\columnwidth]{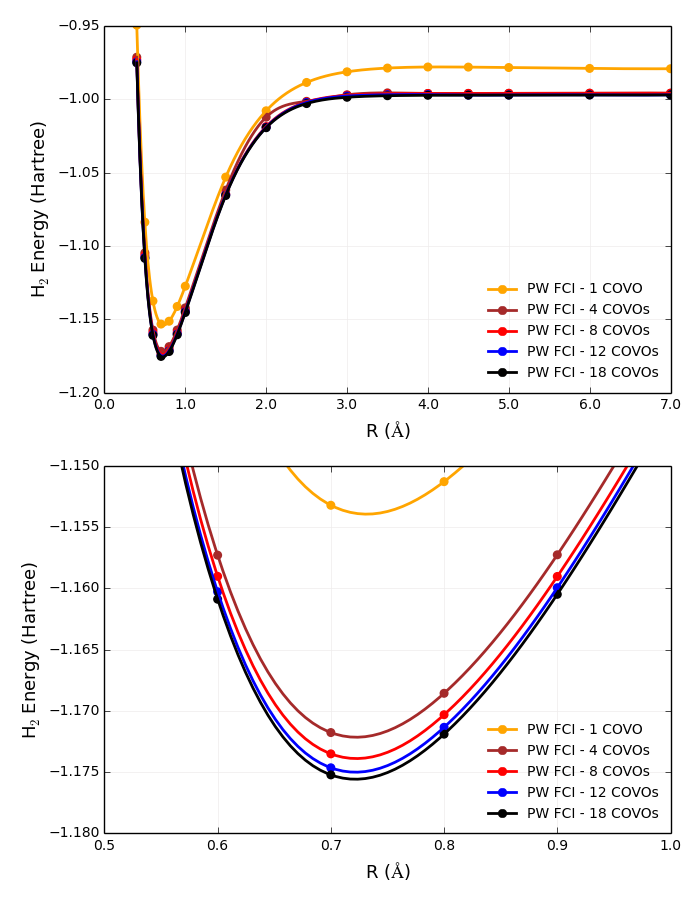}
	\caption{Plots of total energies as a function of distance from plane-wave FCI calculations for the H$_2$ molecule with 1, 4, 8, 12, and 18 correlation optimized virtual orbitals. Top plot shows energy from $R$=0.4 \mbox{\normalfont\AA} to $R$=7.0 \mbox{\normalfont\AA}, and the bottom plot zooms in near the energy minima.}
	\label{fig:h2-pw-fci}
\end{figure}
%%%%%%%%%%%%%%%%%%%%%%%%%%%%%%%%%%%%%%%%%%%%%%%%%%%%%%%%%%%%%%%%
The results for PW FCI calculations of H$_2$ with 1, 4, 8, 12, and 18 COVOs are shown in Fig.~\ref{fig:h2-pw-fci} and Table~\ref{tab:pwfci}.  The average difference error from 18 COVOs is 11.8 kcal/mol, 1.4 kcal/mol, 0.9 kcal/mol, 0.3 kcal/mol for 1, 4, 8, and 12 COVOs calculations respectively. While the error is significant for 1 virtual, the difference is quite small by 4 virtual orbitals, and the error steadily decreases as the number of virtual orbitals is increased. The error seen in the 4 optimized virtual orbital calculation is similar to the 1.6 kcal/mol error seen in the DUCC calculations for the 19 $H_1$ virtual orbitals calculations in section~\ref{sec:h2hfvirt}.  Another measure of the error is the extensivity error.  The energy for large $R$ should be the same as the energy of twice the energy of an isolated H atom.  For the pseudopotential plane-wave method being used the energy of 
2 H atoms is -0.997765 Hartrees (E(H$_1$)=-0.498825 Hartrees). This difference at {$R$=7 \mbox{\normalfont\AA}}  is found to be 11.6 kcal/mol, 1.2 kcal/mol, 1.1 kcal/mol, 0.5 kcal/mol, and 0.4 kcal/mol for 1, 4, 8, 12, and 18 optimized virtual orbital calculations respectively.

Overall the correlation energy at the minimum was found to be -0.035 Hartrees with 4 optimized virtual orbitals, which is comparable to the -0.034 Hartrees found with CCSD/cc-pVTZ. The correlation energy lowers to -0.039 Hartrees with 18 optimized virtual orbitals,  -0.038 Hartrees (12 orbitals), and -0.037 Hartrees (8 orbitals).

%\vspace{1cm}
\section{Quantum simulations of the $^1\Sigma_{ \MakeLowercase{g} }^{+}$ ground state of the H$_2$ molecule using COVO\MakeLowercase{s}}
\label{sec:H2QCresults}

The previous section provides indisputable evidence for the performance of the proposed virtual orbitals for correlation energy recovery. Besides the possible ramifications in quantum chemistry carried out with classical computers, one immediate application is in the realm of quantum simulations. Because the present quantum hardware has not fully matured, hybrid algorithms that leverage classical resources and restrict the workload delegated to quantum computers, namely state preparation and measurements of highly entangled states, are essential to meaningful quantum computations. The COVOs meet this requirement by decreasing the dimensionality of the problem, i.e., by enabling simulations with fewer qubits.
%Figure 5 %%%%%%%%%%%%%%%%%%%%%%%%%%%%%%%%%%%%%%%%%%%%%%%%%%%%%%
\begin{figure}[!ht]
    \centering
    \includegraphics[width=\columnwidth]{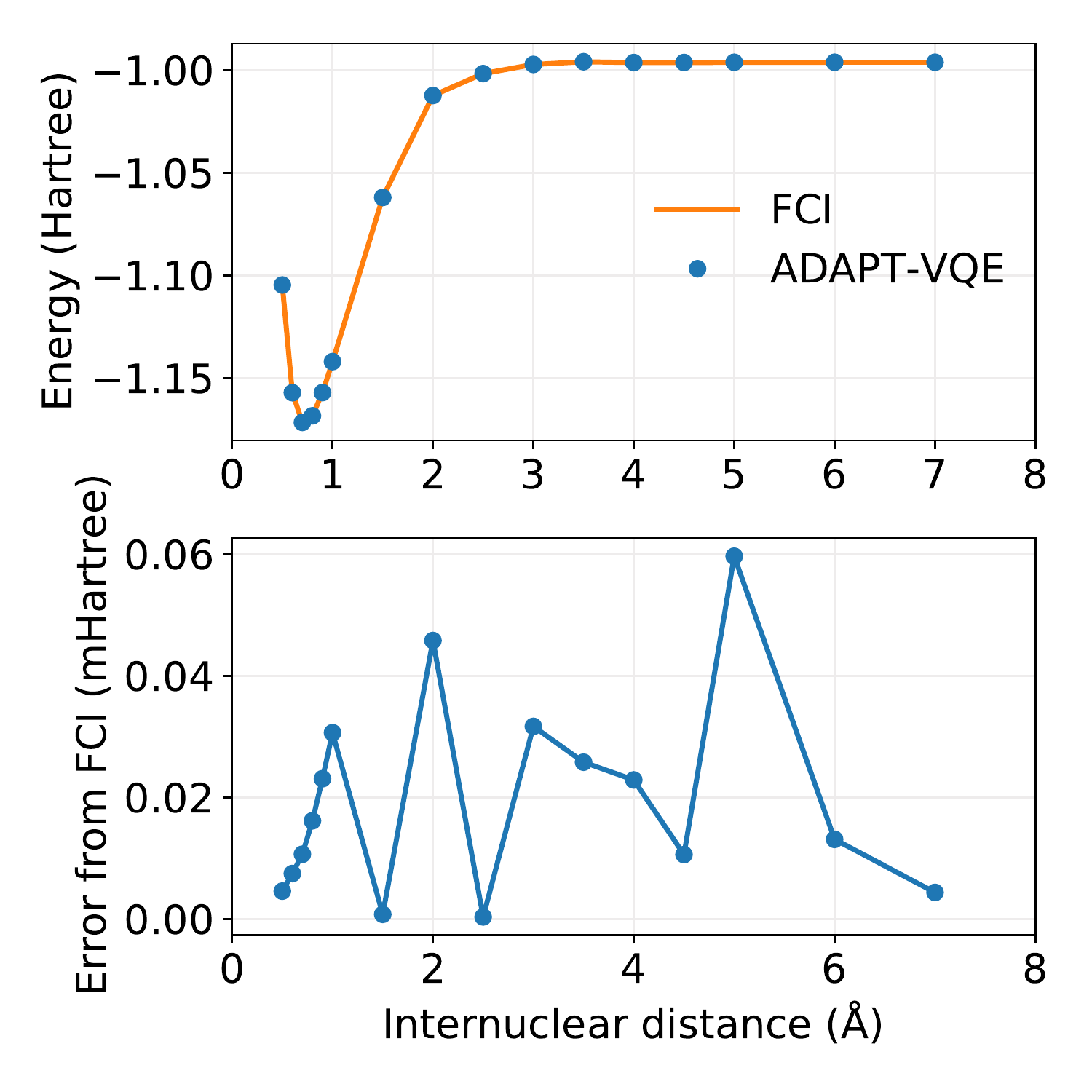}
    \caption{Potential energy curves for FCI and ADAPT-VQE (top) and the deviations in ADAPT-VQE energies with respect to FCI (bottom).}
    \label{fig:adapt_fci_energies}
\end{figure}
%%%%%%%%%%%%%%%%%%%%%%%%%%%%%%%%%%%%%%%%%%%%%%%%%%%%%%%%%%%%%%%%

In order to probe the performance of COVOs in quantum simulations, we use the Hamiltonian with 4 COVOs and simulate the $^1\Sigma_g^{+}$ ground state of H$_2$ in the same bond distances shown in Table~\ref{tab:pwfci} and Figure~\ref{fig:h2-pw-fci}. The ansatz circuit for the simulations is generated according to the ADAPT-VQE algorithm as implemented in the XACC~\cite{xacc1, xacc2} framework for hybrid quantum computing using the tensor network quantum virtual machine (TNQVM) as the noiseless simulator backend~\cite{tnqvm}. In the present study, the ADAPT-VQE cycle is repeated until the norm of the gradient vector falls below $10^{-2}$ and we use an operator pool containing all spin-adapted single and double excitation operators (one- and two-body rotations). A detailed account of ADAPT-VQE is exposed elsewhere~\cite{adapt}. Optimization of the parameterized circuit is conducted with the COBYLA~\cite{cobyla} optimizer as implemented in the \texttt{NLOpt} package~\citeauthor{nlopt}. Results for the simulated potential energy curve are plotted in Figure~\ref{fig:adapt_fci_energies}.

It is evident from Figure~\ref{fig:adapt_fci_energies} that ADAPT-VQE can generate a circuit capable of reproducing the FCI results in the current active space. These simulations deliver a smooth, continuous potential energy curve that tracks the FCI values strikingly well. The deviations from the corresponding FCI energies are all found below $10^{-4}$ mH. This means that not only these simulations deliver results that are well below the conventional chemical accuracy mark, but more importantly in the current context is that this error is inconsequential compared to the effect of noise in case of deployment to actual quantum hardware.
%Figure 6 %%%%%%%%%%%%%%%%%%%%%%%%%%%%%%%%%%%%%%%%%%%%%%%%%%%%%%
\begin{figure}[!ht]
    \centering
    \includegraphics[width=\columnwidth]{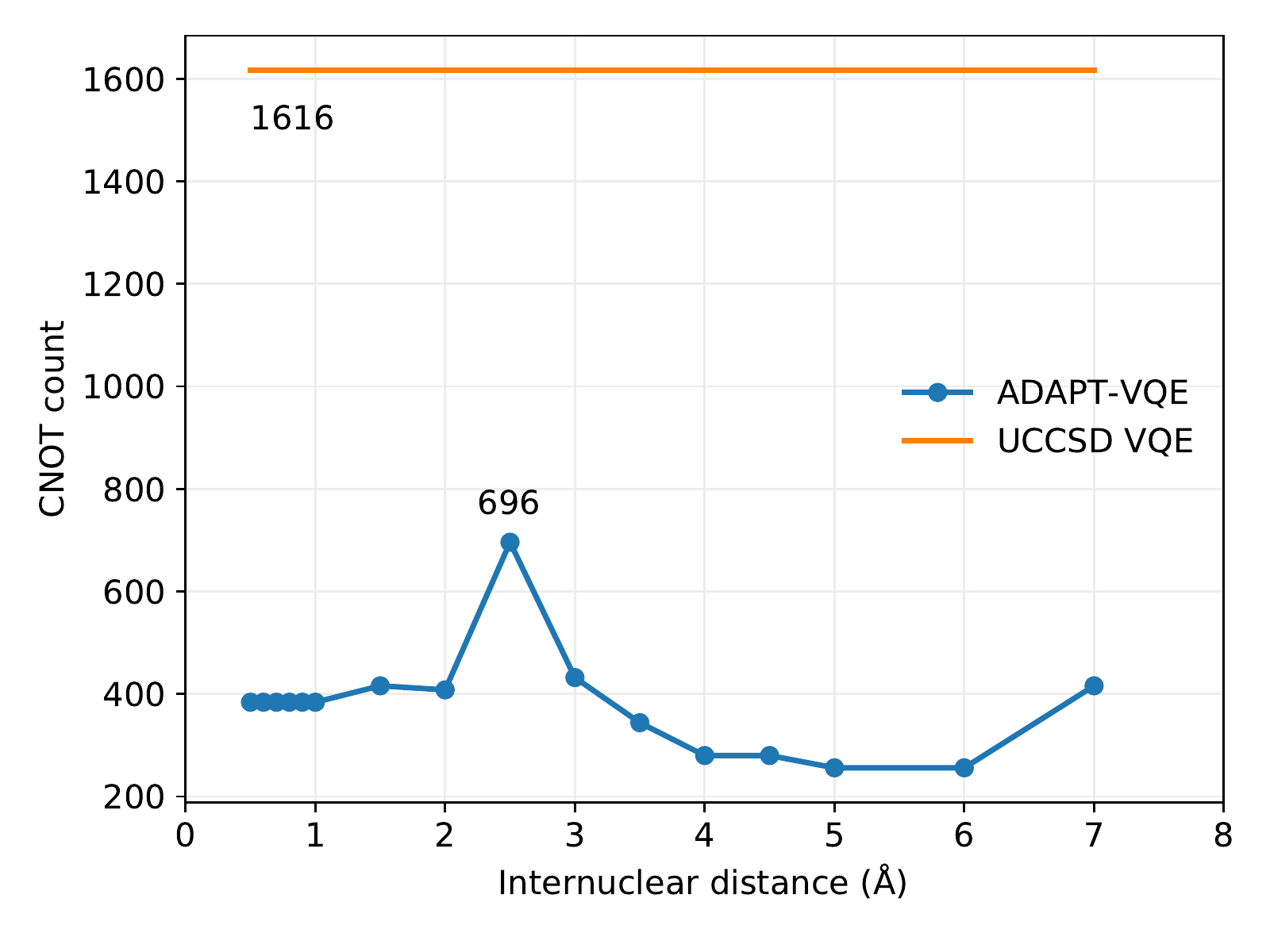}
    \caption{CNOT gate count for ADAPT-VQE and ordinary VQE both with singlet-adapted singles and doubles operators.}
    \label{fig:cnot_count}
\end{figure}
%%%%%%%%%%%%%%%%%%%%%%%%%%%%%%%%%%%%%%%%%%%%%%%%%%%%%%%%%%%%%%%%

It is remarkable that the results degrade little throughout the energy scan, which attests to the aptness and flexibility of ADAPT-VQE in determining an ansatz according to the complexity of the underlying electronic structure. The ansatz in the vicinity of the equilibrium bond length, 0.5-1.0 \mbox{\normalfont\AA} is comprised solely by pair excitations, as would be expected given a restricted HF reference, which means no determinant obtained via one-body rotations can lower the energy below that of HF. As we approach the Coulson-Fischer point~\cite{coulson_fischer} single excitations start to become part of the ansatz, which signals the inadequacy of a restricted reference wave function and that inclusion of these operators enable the ansatz to remain in the $^1\Sigma_g^{+}$ potential energy curve, which means that this flexibility may come at the expense of deeper circuits. Because one-qubit gates tend to be executed in a short timescale and are fairly insensitive to noise, we can use the number of CNOTs present in the circuit as indicative of the complexity in its implementation, which we provide in Figure~\ref{fig:cnot_count}, showing the ansatze generated from ADAPT-VQE are much more affordable than would be obtained by ordinary UCCSD VQE simulations. 

\begin{figure}[!ht]
    \centering
    \includegraphics[width=\columnwidth]{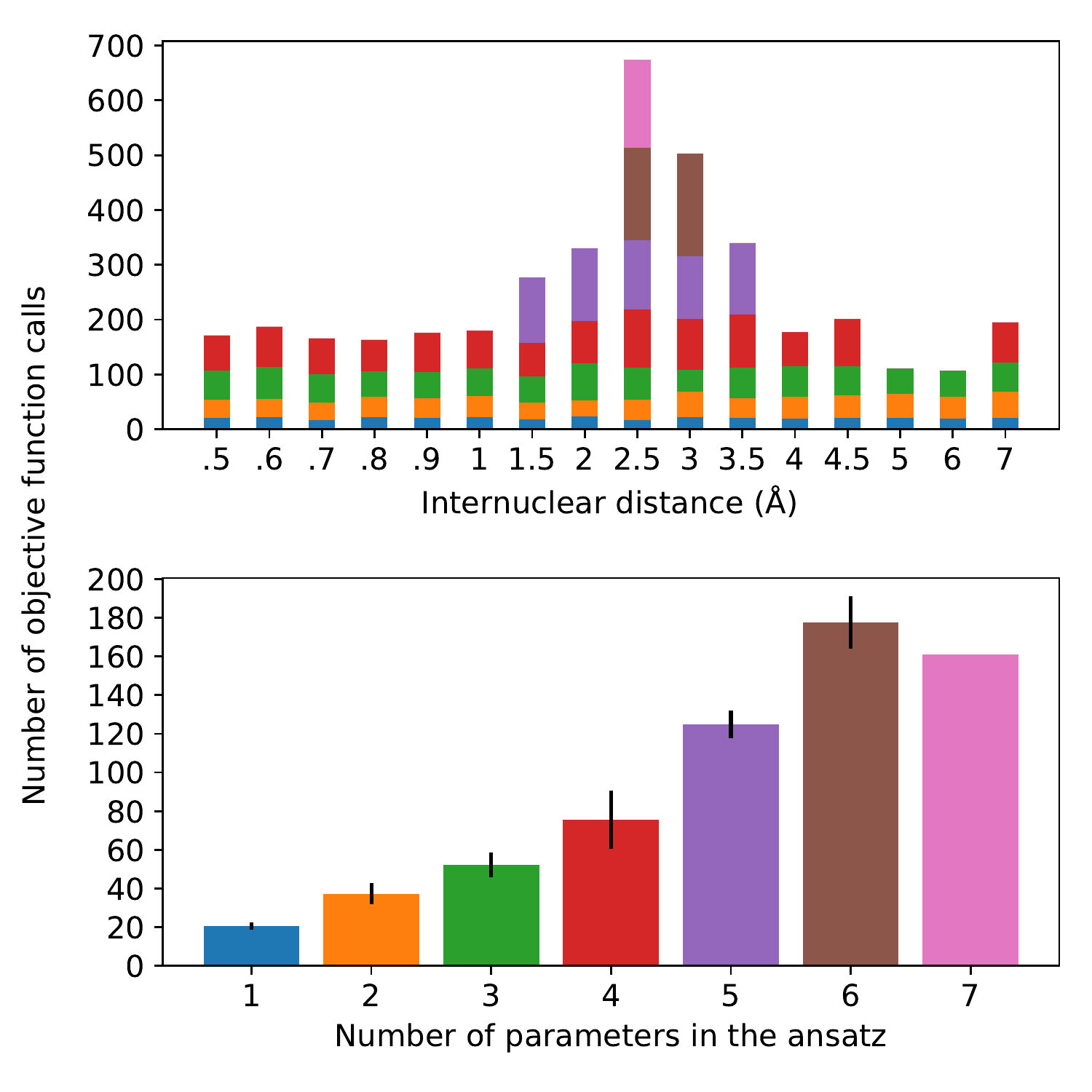}
    \caption{Number of objective function calls as a function of the H-H distance for the different ansatz compositions (top) and average number of objective function calls per ansatz size, with error bars representing one standard devitation (bottom). The bar colors on the top plot represent the ansatz sizes in the bottom plot.}
    \label{fig:optimizer}
\end{figure}

Along these lines, once the operator composition of the ansatz is defined, by virtue of introducing more parameters, we are likely to experience a more arduous optimization of the corresponding parameterized gates. This has a compound effect with the circuit depth since more measurements are needed, each of which requires the circuit to be implemented and measurements to take place. Figure~\ref{fig:optimizer} gives a profile of the optimization performance along the potential energy scan.
%Figure 7 %%%%%%%%%%%%%%%%%%%%%%%%%%%%%%%%%%%%%%%%%%%%%%%%%%%%%%
%%%%%%%%%%%%%%%%%%%%%%%%%%%%%%%%%%%%%%%%%%%%%%%%%%%%%%%%%%%%%%%%

It should come as no surprise that the optimization is more difficult in the regime of stronger correlation. This region also demands a more complex ansatz, as the top plot in Figure~\ref{fig:optimizer} shows that only in this vicinity (1.5 - 3.5 \AA) we observe ansatze with more than four operators. Interestingly, the number of objective function calls do not show large deviations for ansatze with 1-3 operators, regardless of where they are found in the potential energy curve, which is further corroborated by the relatively small errors bars in the corresponding columns of the bottom plot. This observation does not hold as more parameters/operators are introduced in the ansatz in order to accommodate a more complex electronic structure. Thus, with four parameters, not only more calls to the objective function are needed, but there is a more pronounced standard deviation. Anasatze with five or more operators can only be found in the (1.5 - 3.5 \AA), as we can see the calls to the objective function coming from them dominate the overall number of optimization cycles. Due to the scarce occurrence of these ansatze in the current energy scan, the corresponding statistical information that can be derived from these instances is not as reliable. All in all, this plot is valuable in lending additional insight into the resources required to perform these simulations. It is important to mention that for every new ansatz, the variational parameters are initialized at zero. Alternatively, the parameters corresponding to the previously optimizer ansatze could be initialized at their optimal values and the new parameter would be introduced in the ansatz, which would accelerate convergence. Moreover, the convergence profile likely displays pronounced dependence on the chosen optimizer, which is not pursued here.

\section{Conclusion}
\label{sec:conclusions}

%Figure 8 %%%%%%%%%%%%%%%%%%%%%%%%%%%%%%%%%%%%%%%%%%%%%%%%%%%%%%%%
\begin{figure*}[!ht]
    \centering
	\includegraphics[width=\columnwidth]{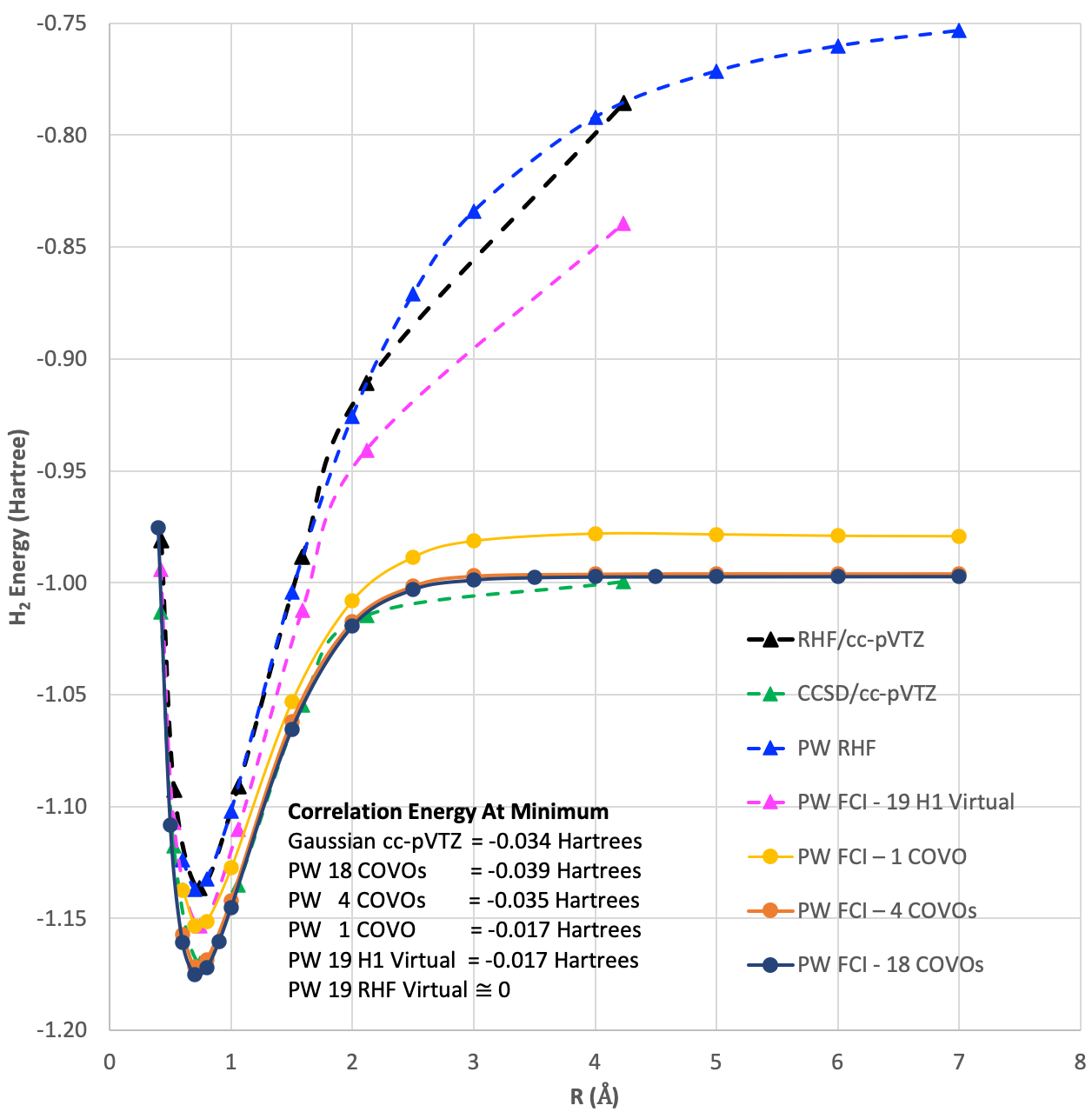}
	\caption{Summary of various of plane-wave and Gaussian basis set RHF and FCI calculations for the $^1\Sigma_g^{+}$  ground state of H$_2$ molecule.
	FCI calculations (not shown) using up to 20 RHF virtual orbitals produced only a negligible amount correlation energy (<1.0e-4 Hartree, i.e. visually the same as RHF results).  It should be noted that for the a two electron H$_2$ molecule CCSD gives the same answer as FCI.}
	\label{fig:Overview}
\end{figure*}
%%%%%%%%%%%%%%%%%%%%%%%%%%%%%%%%%%%%%%%%%%%%%%%%%%%%%%%%%%%%%%%%%%%
In summary, we have developed a new approach for defining virtual spaces with a pseudopotential plane-wave code for use in many-body methods described by second-quantized Hamiltonians. The method is based on optimizing the virtual orbitals to minimize a small select CI Hamiltonian (i.e. COVOs) that contains configurations containing filled RHF orbitals and the one virtual orbital to be optimized.  Subsequent virtual orbitals are optimized in the same way, but with the added constraint of being orthogonal to the previously calculated filled and virtual orbitals.  The method was applied to the simple, but non-trivial, H$_2$ molecule.  As summarized in Figure~\ref{fig:Overview}, these new types virtual orbitals were significantly better at capturing correlation in plane-wave calculations then from virtual spaces from Hartree–Fock and one-electron Hamiltonian, and moreover we were able to obtain good agreement with Gaussian cc-pVTZ basis set results with just 4 virtual orbitals for the H$_2$ molecule. Subsequent calculations showed that the correlation energy converged steadily as more virtual orbitals were included in the calculation.  With 18 virtual orbitals the correlation energies were found to be converged to less than 0.5 kcal/mol.  The robustness of the proposed basis sets is corroborated by its ready applicability to quantum simulations, which in the case of ADAPT-VQE, show remarkable agreement with the classical, exact diagonalization result (FCI) in the same basis set (4 COVOs). 

We are optimistic that these correlation optimized virtual orbitals open up the door to many-body calculations using pseudopotential plane-wave calculations, including coupled cluster, M{\o}ller--Plesset, and Green's function theories as well as other FCI-approaching methods for quantum computers. Future work will focus on using this approach on larger molecular and periodic systems. With the validation granted by our quantum simulations, further studies are called for, including the extension to active-space DUCC downfolded Hamiltonians and work in conjunction with VQE methods.

\section*{Acknowledgments}
This material is based upon work supported by the U.S. Department of Energy (DOE), Office of Science, Office of Basic Energy Sciences, Chemical Sciences, Geosciences, and Biosciences Division through its ``Embedding Quantum Computing into Many-body Frameworks for Strongly Correlated  Molecular and Materials Systems'' project  at Pacific Northwest National Laboratory (PNNL). We also would like to thank the DOE BES Chem CCS, DOE BES Geochemistry and DOE Advanced Scientific Computing Research (ASCR) ECP NWChemEx programs for their support of software development for high-performance computers and computer time needed to carry out the work. PNNL is operated for the U.S. Department of Energy by the Battelle Memorial Institute under Contract DE-AC06-76RLO-1830. This research was also partially supported, thru their support of software development for high-performance computers, by DOE BES Chem CCS, and DOE BES Geosciences programs, as well as the the Exascale Computing Project (17-SC-20-SC), a collaborative effort of the U.S. Department of Energy Office of Science and the National Nuclear Security Administration.  Calculations have been performed using computational resources  at the Pacific Northwest National Laboratory (PNNL).

This research used resources of the National Energy Research Scientific Computing Center (NERSC), a User Facility supported by the Office of Science of the U.S. DOE under Contract No. DE-AC02-05CH11231, and the Argonne ALCF computing center through their early science program. 

%% ORNL and OLCF ack statements
This research also used resources of the Oak Ridge Leadership Computing Facility, which is a DOE Office of Science User Facility supported under Contract DE-AC05-00OR22725
This manuscript has been authored in part by UT-Battelle, LLC under Contract No. DE-AC05-00OR22725 with the U.S. Department of Energy. 
%% ORNL copyright statement
The United States Government retains and the publisher, by accepting the article for publication, acknowledges that the United States Government retains a non-exclusive, paid-up, irrevocable, world-wide license to publish or reproduce the published form of this manuscript, or allow others to do so, for United States Government purposes. The Department of Energy will provide public access to these results of federally sponsored research in accordance with the DOE Public Access Plan.

We would like to thank the NWChem project team and the people that have helped the progress of the NWChem software over the years.  D.C. would like to thank Alexander McCaskey for discussions and help with the software engineering of the quantum algorithm used in this paper.

\section*{Data Availability}
The data sets containing the one and two electron integrals based on COVOs that were used for the VQE quantum computing calculations and the output results from  the ADAPT-VQE calculations can be found in the following github repository,
\small{https://github.com/ebylaska/PWH2-Data/archive/master.zip}

%% Bibtex references
\bibliographystyle{apsrev4-1}
\bibliography{Manuscript}

\end{document}